\pdfoutput=1
\documentclass[12pt]{article}

\font\grande=cmr9.5 scaled \magstep4
\font\medio=cmr9.5 scaled \magstep2
\outer\def\beginsection#1\par{\medbreak\bigskip
      \message{#1}\leftline{\bf#1}\nobreak\medskip
\vskip-\parskip
      \noindent}
\usepackage{graphicx} 
\begin{document}
\bibliographystyle {unsrt}

\titlepage

\begin{flushright}
CERN-PH-TH/2015-161
\end{flushright}

\vspace{10mm}
\begin{center}
{\grande The refractive index of relic gravitons}\\
\vspace{1.5cm}
 Massimo Giovannini
 \footnote{Electronic address: massimo.giovannini@cern.ch}\\
\vspace{1cm}
{{\sl Department of Physics, 
Theory Division, CERN, 1211 Geneva 23, Switzerland }}\\
\vspace{0.5cm}
{{\sl INFN, Section of Milan-Bicocca, 20126 Milan, Italy}}
\vspace*{0.5cm}
\end{center}

\vskip 0.5cm
\centerline{\medio  Abstract}
The dynamical evolution of the refractive index of the tensor modes of the geometry 
produces a specific class of power spectra characterized by a blue (i.e. slightly increasing) slope
which is directly determined by the competition of the slow-roll parameter and of the 
rate of variation of the refractive index. Throughout the conventional stages of the inflationary and post-inflationary evolution, 
the microwave background anisotropies measurements, the 
pulsar timing limits and the big-bang nucleosythesis constraints set stringent bounds
on the refractive index and on its rate of variation. 
Within the physically allowed region of the parameter space the cosmic background of relic gravitons 
leads to a potentially large signal for the ground based detectors (in their advanced version) and for the proposed 
space-borne interferometers. Conversely, the lack of direct detection of the signal will set a qualitatively new bound 
on the dynamical variation of the refractive index. 
\vskip 0.5cm

\noindent

\vspace{5mm}

\vfill
\newpage
\renewcommand{\theequation}{1.\arabic{equation}}
\setcounter{equation}{0}
\section{Introduction}
\label{sec1}
It has been speculated long ago that gravitational waves 
might acquire an effective refractive index when they evolve in curved space-times \cite{sz1}. 
As electromagnetic waves develop a refractive index when they travel in globally 
neutral (but intrinsically charged) media, a similar possibility can also be envisaged 
in the case of linearized gravity. In this investigation it is suggested that the consistent variation 
of the refractive index throughout the conventional stages of the cosmological evolution 
leads to the production of a stochastic background of relic gravitons with blue spectral slopes.

Relic gravitons are known to be produced in the early Universe thanks to the pumping action of the gravitational field \cite{gr}.
This phenomenon occurs in a variety of different scenarios and, in particular, in the case of conventional inflationary models (see e.g. 
\cite{rub,FS1,max01,paul,TR1,TR2,TR3,max12} for an incomplete but potentially interesting list of time-ordered references). 
While inflationary models typically predict 
decreasing slopes, in the conventional lore blue spectral indices of the relic graviton backgrounds can arise when a long 
phase (dominated by stiff sources) takes place after inflation but prior to the dominance of radiation \cite{max01}.  Other less conventional possibilities include gravity theories which are not of Einstein-Hilbert type (see e.g. last paper of Ref. \cite{rub}) and the violation of the dominant energy condition in the early Universe \cite{doc}. In this paper we are going to argue that blue spectral slopes may arise from a comparatively more mundane possibility, namely the temporal variation of the refractive index of the tensor modes while the evolution of the background geometry follows exactly the same patterns of the concordance paradigm.

The stochastic backgrounds of relic gravitons are subjected to three complementary classes of constraints. The first class of direct limits stems from the temperature and polarization anisotropies 
of the cosmic microwave background \cite{cmb1,cmb2,cmb3} and it is customarily expressed in terms of a bound on the tensor to scalar ratio $r_{T}(k_{p})$ at a conventional pivot wavenumber $k_{p}$ where the large-scale power spectra are assigned. 
The pulsar timing measurements \cite{pulsar1,pulsar2} impose instead an upper bound on the cosmic graviton background at a typical frequency roughly 
corresponding  to the inverse of the observation time along which the pulsars timing has been monitored.
Finally the big-bang nucleosynthesis limits \cite{bbn1} set an indirect  constraint  
on the extra-relativistic species (and, among others, on the relic gravitons)  at the time when light nuclei have been firstly formed.

In this paper we intend to compute the cosmic background of the relic gravitons induced by the consistent variation 
of the refractive index during the early stages of the evolution of the geometry. The appropriately constrained spectra shall be compared with the frequency window of the ground-based interferometers such as Ligo/Virgo \cite{virgoligo1,virgoligo2},  Geo600
\cite{GEO} and the recently proposed Kagra \cite{kagra} (ideal prosecution of the Tama300 experiment \cite{TAMA}).  There also exist daring projects of wide-band detectors in space like the Lisa interferometer \cite{lisa} (in one of its different incarnations)  or  the Bbo/Decigo \cite{BBODECIGO} project\footnote{ The acronyms appearing in this and in the previous sentences refer to the corresponding projects: Lisa (Laser Interferometer Space Antenna), Bbo (Big Bang Observer), Decigo (Deci-hertz Interferometer Gravitational Wave Observatory) and Kagra (Kamioka Gravitational Wave Detector).}. 

The variation of the refractive index along the different stages of the evolution of the background must be 
continuous and differentiable at least once. This basic requirement stems directly from the evolution 
equations of the tensor modes of the geometry. In the case of a conventional inflationary and post-inflationary evolution the scale factor approximately evolves as\footnote{We are assuming here a conformally flat 
Friedmann-Robertson-Walker background metric $\overline{g}_{\mu\nu} = a^2(\tau) \eta_{\mu\nu}$ where $a(\tau)$ is the scale factor, $\tau$ denotes the conformal time coordinate and $\eta_{\mu\nu}$ is the Minkowski metric. This is the simplest way of complying with the concordance scenario 
where the extrinsic curvature is always much larger than the intrinsic (spatial) curvature.} 
\begin{eqnarray}
a_{inf}(\tau) &=& \biggl(- \frac{\tau_{1}}{\tau}\biggr)^{\beta}, \qquad \tau \leq - \tau_{1},
\label{sfinf}\\
a_{r}(\tau) &=& \frac{\beta \tau + (\beta+1) \tau_{1}}{\tau_{1}}, \qquad -\tau_{1} < \tau \leq \tau_{2},
\label{sfrad}\\
a_{m}(\tau) &=& \frac{[ \beta \tau + \beta \tau_{2} + 2 (\beta+1) \tau_{1}]^2}{4 \tau_{1} [ \beta \tau_{2} + (\beta+ 1)\tau_{1}]}, \qquad \tau > \tau_{2},
\label{sfmat}
\end{eqnarray}
where $\tau_{1}$ coincides with the end of the inflationary phase and $\tau_{2}$ coincides with the time of matter-radiation equality; note that $\beta\to 1$ 
in the case of a pure de Sitter phase and $\beta = 1 - {\mathcal O}(\epsilon)$ in the quasi-de Sitter case.
The transition to the domination of the dark energy will be discussed later on since, in practice, it does not affect the slope and it has a mild 
effect on the amplitude of the spectrum. Equations (\ref{sfinf}), (\ref{sfrad}) and (\ref{sfmat}) are all continuous with their first derivatives at the transition points\footnote{This means, more specifically, that for  $\tau =-\tau_{1}$ we have that  $a_{i}(-\tau_{1}) = a_{r}(-\tau_{1})$ and $a_{i}'(-\tau_{1}) = a_{r}'(-\tau_{1})$. Similarly at the second transition point $a_{r}(\tau_{2}) = a_{m}(\tau_{2})$ and $a_{r}'(\tau_{2}) = a_{m}'(\tau_{2})$. }.  

Since relic gravitons are produced because of the pumping action of the background curvature (containing second derivatives of the scale factor), the continuity of  Eqs. (\ref{sfinf}), (\ref{sfrad}) and (\ref{sfmat}) at the transition points ensures that the evolution equations of the relic gravitons will not have singularities in $\tau_{1}$ and $\tau_{2}$ but, at most, jump discontinuities. To guarantee a continuous variation of the refractive index without imposing further conditions, we are led to the following parametrization:
\begin{equation}
n(\tau) = n_{1} a^{\alpha}(\tau), \qquad \alpha >0,
\label{opt1}
\end{equation}
where $\alpha$ measures, in practice, the rate of variation of the refractive index in units of the Hubble rate. 
Equation (\ref{opt1}) implies that $n(\tau)$ is automatically continuous and differentiable in $\tau_{1}$ and $\tau_{2}$ provided the scale factor 
shares the same properties at the transition points. The parametrization (\ref{opt1}) is minimal insofar as it contains only two arbitrary parameters, namely $n_{1}$ and $\alpha$. Furthermore the value of $n_{1}$ is not totally arbitrary\footnote{For practical reasons the bounds on the amplitude of the spectral index 
can be more simply expressed in terms of $n_{i} = n(\tau_{i})$ where $\tau_{i}$ coincides with beginning of the inflationary phase. Since 
the evolution during inflation is known $n_{1}$ can be easily related to $n_{i}$ by a simple redshift factor.} since at the onset of the  inflationary phase the 
refractive index must be larger than (or equal to) $1$ to avoid superluminal phase and group velocities. Note that when the background is ever expanding, the positivity of $\alpha$ guarantees that this condition is preserved throughout the evolution of the geometry. 

The plan of this paper is the following. In section \ref{sec2} the most relevant technical aspects of the analysis are derived.
The power spectra and the spectral energy density of the relic gravitons 
are computed in section \ref{sec3} and in the framework of the conventional cosmological evolution. 
In section \ref{sec4} the spectra of the relic gravitons are confronted with all the available constraints. In the final part of section \ref{sec4} the prospects for the wide-band detectors of gravitational waves are illustrated. Section \ref{sec5} contains our concluding remarks.  Some useful but lengthy results are collected in appendices \ref{APPA} and \ref{APPB}. 

\renewcommand{\theequation}{2.\arabic{equation}}
\setcounter{equation}{0}
\section{Relic gravitons with refractive index}
\label{sec2}
\subsection{Basic definitions}
The two polarizations of the gravitational wave are defined as
 \begin{equation}
 e_{ij}^{(\oplus)}(\hat{k}) = (\hat{m}_{i} \hat{m}_{j} - \hat{q}_{i} \hat{q}_{j}), \qquad 
 e_{ij}^{(\otimes)}(\hat{k}) = (\hat{m}_{i} \hat{q}_{j} + \hat{q}_{i} \hat{m}_{j}),
 \label{ST0}
 \end{equation}
where  $\hat{k}_{i} = k_{i}/|\vec{k}|$,  $\hat{m}_{i} = m_{i}/|\vec{m}|$ and $\hat{q} =q_{i}/|\vec{q}|$ are three mutually 
orthogonal directions  and $\hat{k}$ is oriented along the direction of propagation of the wave. It follows directly from Eq. (\ref{ST0}) that $e_{ij}^{(\lambda)}\,e_{ij}^{(\lambda')} = 2 \delta_{\lambda\lambda'}$ while the sum over the polarizations gives:
\begin{equation}
\sum_{\lambda} e_{ij}^{(\lambda)}(\hat{k}) \, e_{m n}^{(\lambda)}(\hat{k}) = \biggl[p_{m i}(\hat{k}) p_{n j}(\hat{k}) + p_{m j}(\hat{k}) p_{n i}(\hat{k}) - p_{i j}(\hat{k}) p_{m n}(\hat{k}) \biggr];
\label{ST0B} 
\end{equation}
where $p_{ij}(\hat{k}) = (\delta_{i j} - \hat{k}_{i} \hat{k}_{j})$. Defining the Fourier transform of $h_{ij}(\vec{x},\tau)$ as 
\begin{equation}
h_{ij}(\vec{x},\tau) = \frac{1}{(2\pi)^{3/2}}\, \sum_{\lambda}  \, \int d^{3} k \, h_{ij}(\vec{k},\tau)\,\, e^{- i \vec{k}\cdot\vec{x}},
\label{St0C}
\end{equation}
the tensor power spectrum ${\mathcal P}_{T}(k,\tau)$ determines the two-point function at equal times:
\begin{equation}
\langle h_{ij}(\vec{k},\tau) \, h_{mn}(\vec{p},\tau) \rangle = \frac{2\pi^2}{k^3} {\mathcal P}_{T}(k,\tau) \, {\mathcal S}_{ijmn}(\hat{k}) \delta^{(3)}(\vec{k} +\vec{p}),
\label{ST1}
\end{equation}
where, ${\mathcal S}_{ijmn}(\hat{k}) = \sum_{\lambda} e_{ij}^{(\lambda)}(\hat{k}) \, e_{m n}^{(\lambda)}(\hat{k})/4$.
The analog of Eq.  (\ref{ST1}) for $h^{\prime}_{ij}(\vec{k},\tau)$ is given by:
\begin{equation}
\langle h^{\prime}_{ij}(\vec{k},\tau) \, h^{\prime}_{mn}(\vec{p},\tau) \rangle = \frac{2\pi^2}{k^3} {\mathcal Q}_{T}(k,\tau) \, {\mathcal S}_{ijmn}(\hat{k}) \delta^{(3)}(\vec{k} +\vec{p}),
\label{ST1D}
\end{equation}
where ${\mathcal Q}_{T}(k,\tau)$ is the corresponding power spectrum and where the prime denotes a derivation with respect to the conformal time coordinate $\tau$.
Note that Eq. (\ref{ST1}) follows the same conventions used when deriving the spectrum of curvature perturbations  on comoving orthogonal hypersurfaces (customarily denoted by ${\mathcal R}(\vec{x},\tau)$) 
\begin{equation}
\langle {\mathcal R}(\vec{k},\tau) \, {\mathcal R}(\vec{p},\tau) \rangle = \frac{2\pi^2}{k^3} {\mathcal P}_{{\mathcal R}}(k,\tau)  \delta^{(3)}(\vec{k} +\vec{p}),
\label{ST1C}
\end{equation}
which is exactly the quantity employed to set the initial conditions for the evolution of the temperature and polarization anisotropies of the Cosmic Microwave Background \cite{cmb1,cmb2,cmb3}. We also remind for future convenience that, according to the standard convention,
the scalar power spectrum is assigned as:
\begin{equation}
{\mathcal P}_{{\mathcal R}}(k)= {\mathcal A}_{{\mathcal R}} \biggl(\frac{k}{k_{p}}\biggr)^{n_{s} -1}, \qquad k_{p} = 0.002\, \mathrm{Mpc}^{-1},
\label{ST1Ca}
\end{equation}
where $k_{p}$ is called the pivot scale, $n_{s}$ is the scalar spectral index and ${\mathcal A}_{R}$ is the amplitude of the scalar power 
spectrum at the pivot scale.  

\subsection{Power spectra and spectral energy density}
The equation obeyed by $h_{ij}(\vec{x},\tau)$ follows from the second order action: 
\begin{equation}
S = \frac{1}{8 \ell_{P}^2} \int d^{3}x \int d\tau \, a^2(\tau) \biggl[ \partial_{\tau} h_{ij} \partial_{\tau} h_{ij} - \frac{1}{n^2(\tau)} \partial_{k} h_{ij} \partial_{k} h_{ij} \biggr],
\label{action1}
\end{equation}
which reduces to the conventional action \cite{max12,FP} in the limit $n(\tau) \to 1$.
Note that in Eq. (\ref{action1}) $\ell_{P} = \sqrt{8 \pi G} = 8\pi/M_{P}$ and $M_{P} = 1.22\times 10^{19} \mathrm{GeV}$. 
From Eq.  (\ref{action1}) the equations of motion for $h_{i}^{j}$ are:
\begin{equation}
h_{ij}^{\prime\prime} + 2 {\mathcal H} h_{ij}^{\prime} - \frac{\nabla^2 h_{ij} }{n^2(\tau)}= 0,
\label{mot1A}
\end{equation}
where ${\mathcal H} = (\ln{a})' = a\,H$ and $H$ is the conventional Hubble rate. In Eq. (\ref{mot1A})  
the contribution of the (transverse and traceless) anisotropic stress has been neglected. At low frequencies and in the concordance 
paradigm the contribution to the anisotropic stress is due to the presence of (effectively massless) neutrinos \cite{wnu1}.  At high frequencies the anisotropic 
stress induced by waterfall fields may lead to an enhancement of the spectral energy density (see third paper in Ref. \cite{max12}). Both effects will be neglected in what follows for two 
independent reasons. We shall neglect neutrinos because they are known to suppress the energy density of the relic gravitons at intermediate 
frequencies but their numerical relevance is not strictly essential for the present considerations. The waterfall field, on the contrary, may lead to large effects which are, however, model dependent, insofar as they arise in a given and specific class of inflationary scenarios. 

In the absence of anisotropic stress $h_{ij}(\vec{x},\tau)$ can be quantized and the corresponding 
field operator is:
\begin{equation}
\hat{h}_{ij}(\vec{x},\tau) = \frac{\sqrt{2} \ell_{P}}{(2\pi)^{3/2}}\sum_{\lambda} \int \, d^{3} k \,\,e^{(\lambda)}_{ij}(\vec{k})\, [ F_{k,\lambda}(\tau) \hat{a}_{\vec{k}\,\lambda } e^{- i \vec{k} \cdot \vec{x}} + F^{*}_{k,\lambda}(\tau) \hat{a}_{\vec{k}\,\lambda }^{\dagger} e^{ i \vec{k} \cdot \vec{x}} ],
\label{T8}
\end{equation}
where $F_{k\,\lambda}(\tau)$ is the (complex) mode function obeying Eq. (\ref{mot1A}) and the sum is performed over the two 
physical polarizations of Eq. (\ref{ST0}); note that $[\hat{a}_{\vec{k},\lambda}, \hat{a}^{\dagger}_{\vec{p}, \lambda^{\prime}} ]= \delta_{\lambda\lambda^{\prime}}\, \delta^{(3)}(\vec{k}- \vec{p})$.  
The same expansion of Eq. (\ref{T8}) can be obtained for the derivative of the amplitude 
\begin{equation}
\hat{h}_{ij}^{\prime}(\vec{x},\tau) = \frac{\sqrt{2} \ell_{P}}{(2\pi)^{3/2}}\sum_{\lambda} \int \, d^{3} k \,\,e^{(\lambda)}_{ij}(\vec{k})\, [ G_{k,\lambda}(\tau) \hat{a}_{\vec{k}\,\lambda } e^{- i \vec{k} \cdot \vec{x}} + G^{*}_{k,\lambda}(\tau) \hat{a}_{\vec{k}\,\lambda }^{\dagger} e^{ i \vec{k} \cdot \vec{x}} ],
\label{T8a}
\end{equation}
where, this time, $G_{k} = F_{k}^{\prime}$. The power spectra introduced in Eqs. (\ref{ST1}) and (\ref{ST1D}) become, in this specific case: 
\begin{eqnarray}
{\mathcal P}_{\mathrm{T}}(k,\tau) &=& \frac{4 \ell_{P}^2\,\, k^3}{\pi^2} |F_{k}(\tau)|^2,
\label{T10b}\\
 {\mathcal Q}_{\mathrm{T}}(k,\tau) &=& \frac{4 \ell_{P}^2\,\, k^3}{\pi^2} |G_{k}(\tau)|^2.
\label{T10bb}
\end{eqnarray}
The mode functions $F_{k}(\tau)$ obey the following equation which is the Fourier space analog of Eq. (\ref{mot1A}):
\begin{equation}
F_{k}^{\prime\prime} + 2 \frac{a^{\prime}}{a} F_{k}^{\prime} + \frac{k^2}{n^2(\tau)} F_{k} =0,
\label{MODEA}
\end{equation}
that can also be written as 
\begin{equation}
f_{k}^{\prime\prime} +\biggl[ \frac{k^2}{n^2(\tau)} - \frac{a^{\prime\prime}}{a} \biggr] f_{k} =0.
\label{MODEB}
\end{equation}
Following Ford and Parker \cite{FP} (see also \cite{Lan,isa,other} for complementary approaches) the energy density of the relic gravitons can be written as
\begin{equation}
\rho_{gw} = \frac{1}{8 \ell_{P}^2 a^2 } \biggl[ \partial_{\tau} h_{ij} \partial_{\tau} h_{ij} + \frac{1}{n^2(\tau)} \partial_{k} h_{ij} \partial_{k}h_{ij} \biggr].
\label{T8d}
\end{equation}
Within the established notations\footnote{We take the opportunity for an elementary observation which is however 
rather crucial to avoid potential confusions: in this paper the natural logarithms will be denoted by ``$\ln$'' while the common logarithms 
will be denoted by ``$\log$''.}  the energy density per logarithmic interval of wavenumber becomes
\begin{equation}
\frac{d \rho_{gw}}{d \ln{k}} = \frac{1}{8 \ell_{P}^2 a^2 } \biggl[ \frac{k^2}{n^2(\tau)} {\mathcal P}_{T}(k, \tau) + {\mathcal Q}_{T}(k,\tau)\biggr] \to \frac{k^2}{4\ell_{P}^2 \,a^2(\tau) \,n^2(\tau)} {\mathcal P}_{T}(k,\tau), 
\label{T8e}
\end{equation}
where the final result holds when the modes are inside the Hubble radius since, in this case, $k^2 {\mathcal P}_{T}(k,\tau)/n^2(\tau) \to {\mathcal Q}_{T}(k,\tau)$. In the opposite limit we have instead that ${\mathcal Q}_{T}(k,\tau) \to {\mathcal H}^2 {\mathcal P}_{T}(k,\tau)$. 
When discussing the graviton spectra over various orders of magnitude in frequency it is more practical to deal with the spectral energy density of the relic gravitons in critical units per logarithmic interval of wavenumber:
\begin{equation}
\Omega_{gw}(k,\tau) = \frac{1}{\rho_{crit}} \frac{d \rho_{gw}}{d \ln{k}}, \qquad \rho_{crit} = 3 H^2/\ell_{P}^2.
\label{T12}
\end{equation}
The energy density of the relic gravitons per logarithmic interval of comoving wavenumber (or logarithmic interval of comoving 
frequency) introduced in Eqs. (\ref{T8e}) and (\ref{T12}) will be occasionally called spectral energy density of the cosmic 
graviton background. 

\subsection{Practical time parametrizations}

We conclude this section with few remarks involving the time parametrizations. As we saw the evolution of the mode functions can be perfectly well discussed in the conformal time parametrization. However, for an explicit solution of the equations, it is convenient to use the $\eta$-time parametrization. Indeed, the action can be expressed in a simpler form by introducing a different time coordinate defined by $d \tau = n(\eta) d\eta$. In this case the action of Eq. (\ref{action1})
can be expressed as: 
\begin{equation}
S = \frac{1}{8 \ell_{P}^2} \int d^{3}x \int d\eta \, b^2(\eta) \biggl[ \partial_{\eta} h_{ij} \partial_{\eta} h_{ij} - \partial_{k} h_{ij} \partial_{k} h_{ij} \biggr], \qquad 
b(\eta) = \frac{a(\eta)}{\sqrt{n(\eta)}}.
\label{action2}
\end{equation}
The mode expansion is analog to Eq. (\ref{T8}) and it is given by: 
\begin{equation}
\hat{h}_{ij}(\vec{x},\eta) = \frac{\sqrt{2} \ell_{P}}{(2\pi)^{3/2}}\sum_{\lambda} \int \, d^{3} k \,\,e^{(\lambda)}_{ij}(\vec{k})\, [ \overline{F}_{k,\lambda}(\eta) \hat{a}_{\vec{k}\,\lambda } e^{- i \vec{k} \cdot \vec{x}} + \overline{F}^{*}_{k,\lambda}(\eta) \hat{a}_{\vec{k}\,\lambda }^{\dagger} e^{ i \vec{k} \cdot \vec{x}} ],
\label{T8aa}
\end{equation}
where, however, the evolution equation obeyed by $\overline{F}_{k}(\eta)$ differs from Eq. (\ref{MODEA}) and it is given by 
\begin{equation}
\frac{\partial^2 \overline{F}_{k}}{\partial\eta^2} + \frac{2}{b} \biggl(\frac{\partial b}{\partial\eta}\biggr) \frac{\partial \overline{F}_{k}}{\partial \eta} + k^2 \overline{F}_{k} =0.
\label{mode2}
\end{equation}
The evolution of the mode function rescaled through $b(\eta)$ will then read
\begin{equation}
\frac{\partial^2 \overline{f}_{k}}{\partial\eta^2} + \biggl[ k^2 - \frac{1}{b}\biggl(\frac{\partial^2 b}{\partial\eta^2}\biggr)\biggr] \overline{f}_{k}=0, \qquad \overline{f}_{k} = b(\eta) \overline{F}_{k}(\eta). 
\label{mode3}
\end{equation}
The parametrization of Eq. (\ref{opt1}) 
implies that power-law behaviours in the $\tau$-parametrization translate into power-laws in the $\eta$-parametrization. This is always 
true except for the case when the relation between $\eta$ and $\tau$ is logarithmic. This happens, for instance, when $\alpha = 1$ and the scale factor evolves 
during the radiation-dominated phase (i.e. Eq. (\ref{sfrad})). The same thing happens when $\alpha =1/2$ and the scale factor is the one 
of dusty matter (as in Eq. (\ref{sfmat})). Recalling  Eq. (\ref{action2}) for the definition of $b(\eta)$, if $\alpha=1$ we have that $b(\eta)\propto \eta$ during the radiation epoch; similarly when $\alpha = 1/2$ we also have that $b(\eta)\propto \eta$ during the matter phase. In these two cases Eq. (\ref{mode3}) has a plane-wave solution in $\eta$. Even if the cases  $\alpha =1$ and $\alpha=1/2$ must be separately treated, the results do not have a prominent  physical meaning since they belong to a region of the 
parameter space which is anyway phenomenologically excluded. We shall therefore proceed in the discussion by assuming, for the sake of conciseness, that $\alpha \neq 1$ and $\alpha \neq 1/2$.

\renewcommand{\theequation}{3.\arabic{equation}}
\setcounter{equation}{0}
\section{Power spectra in the different phases}
\label{sec3}
The evolution of the mode functions of Eqs. (\ref{MODEA}) and (\ref{MODEB}) must 
be solved by taking into account the evolution of the refractive index (see Eq. (\ref{opt1})) in each of the different stages defined, respectively, by Eqs. (\ref{sfinf}), (\ref{sfrad}) and (\ref{sfmat}).  At the practical level the strategy is to pass from the $\tau$-parametrization to the 
$\eta$-parametrization and then transform back the obtained result in the conformal time coordinate.  
Since this procedure is algebraically lengthy but completely straightforward, we shall simply present the final result for the correctly normalized mode function and avoid pedantic details. We finally mention that it is useful to introduce, in some of the forthcoming equations, the obvious notation 
\begin{equation}
\omega_{inf}(\tau) = \frac{k}{n_{1} a_{inf}^{\alpha}(\tau)}, \qquad \omega_{r}(\tau) = \frac{k}{n_{1} a_{r}^{\alpha}(\tau)}, \qquad \omega_{m}(\tau) = \frac{k}{n_{1} a_{m}^{\alpha}(\tau)},
\label{opt2}
\end{equation}
where the scale factors in the different epochs are parametrized as in Eq. (\ref{sfinf})--(\ref{sfmat}).

\subsection{Power spectrum during inflation}
When the scale factor and the refractive index are given, respectively, by Eqs. (\ref{sfinf}) and (\ref{opt1}),  the normalized solution of Eq. (\ref{MODEB})  is given by:
\begin{equation}
f_{k}(\tau)= \frac{{\mathcal D}_{i}}{\sqrt{ 2 \omega_{inf}(\tau)}} \,\, \sqrt{- \omega_{inf}(\tau)\, \tau} \,\, H^{(1)}_{\mu}[g_{i}(\tau)], \qquad \mu= \frac{3 - \epsilon}{2 ( 1 - \epsilon) |1 + \alpha \beta|},
\label{inf1} 
\end{equation}
where $H_{\mu}^{(1)}[g_{i}(\tau)]$ is the Hankel function of the first kind \cite{abr,tric} with argument $g_{i}(\tau)$ and index $\mu$. Equation (\ref{inf1}) has been derived in the 
case where the slow-roll parameter $\epsilon = - \dot{H}/H^2$ is constant in time. In this case it turns out that $\beta = 1/(1 - \epsilon)$ since 
$a H= - 1/[(1 -\epsilon)\tau]$.  In Eq.  (\ref{inf1}) the normalization $|{\mathcal D}_{i}| = \sqrt{\pi/(2 | 1 + \alpha\beta|)}$ 
guarantees that,  up to an irrelevant phase, Eq. (\ref{inf1})  coincides with a 
plane wave in the large argument limit of Hankel functions. Since $\omega_{inf}(\tau)$ depends on $\tau$
 it is practical to introduce a single argument $g_{i}(\tau)$ as\footnote{Since $\alpha \geq 0$ and $ \beta = 1/(1 - \epsilon)$ the absolute value is pleonastic. During the radiation and matter epochs the analog factors are not necessarily positive definite. To keep a homogeneous 
 notation the absolute values have been always included even when not mandatory.}
\begin{equation}
g_{i}(\tau) = - \frac{\tau\, \omega_{inf}(\tau)}{| 1 + \alpha \beta|}= \frac{k \tau_{1}}{| 1 + \alpha \beta| \, n_{1}} \,  \biggl(- \frac{\tau}{\tau_{1}} \biggr)^{ 1 + \alpha \beta}.
\label{inf2}
\end{equation}

Inserting Eq. (\ref{inf1}) into Eq. (\ref{T10b}) we obtain, after some 
algebra, the explicit expression of the inflationary power spectrum:
\begin{equation}
{\mathcal P}_{T}(k,\tau) = 8 \biggl(\frac{H_{1}}{M_{P}}\biggr)^2 \frac{|k\,\tau_{1}|^3}{\pi |1 + \alpha\beta|}\,\, \biggl(-\frac{\tau}{\tau_{1}}\biggr)^{ 1 + 2\beta}\,\,\biggl| H_{\mu}^{(1)}[ g_{i}(\tau)]\biggr|^2,
\label{inf3}
\end{equation}
where $\mu$ can also be expressed as $\mu = (3-\epsilon)/[ 2 ( 1 - \epsilon + \alpha)]$. Using Eq. (\ref{inf2}) and considering the modes that are larger than the Hubble radius, Eq. (\ref{inf3}) becomes:
\begin{eqnarray}
{\mathcal P}_{T}(k, k_{max}) &=& {\mathcal C}(\epsilon, \alpha) \,\, n_{1}^{3 - n_{T}(\epsilon,\alpha)}\,\,\biggl(\frac{H_{1}}{M_{P}}\biggr)^2 \biggl(\frac{k}{k_{max}}\biggr)^{n_{T}(\epsilon,\alpha)},
\nonumber\\
{\mathcal C}(\epsilon, \alpha) &=& \frac{2^{ 6- n_{T}(\epsilon,\alpha)}}{\pi^2} \Gamma^2\biggl[\frac{3 - n_{T}(\epsilon,\alpha)}{2}\biggr] \biggl| 1 + \frac{\alpha}{1 - \epsilon}\biggr|^{ 2 - n_{T}(\epsilon,\alpha)}, 
\nonumber\\
n_{T}(\epsilon,\alpha) &=& 3 - \frac{3 - \epsilon}{(1 - \epsilon + \alpha)}.
\label{inf5}
\end{eqnarray}
Denoting with $\Omega_{{R}0}$ the present value of the critical fraction of radiative species (in the concordance paradigm 
photons and neutrinos) and with  ${\mathcal A}_{R}$ the amplitude of the scalar power spectrum at the pivot scale (see Eq. (\ref{ST1Ca}))
the value of $k_{max}$ can be expressed, for instance, in $\mathrm{Mpc}^{-1}$ units:
\begin{equation}
\biggl(\frac{k_{max}}{\mathrm{Mpc^{-1}}} \biggr) = 2.247 \times 10^{23}\, \biggl(\frac{H_{r}}{H_{1}}\biggr)^{\gamma -1/2} \biggl(\frac{\epsilon}{0.01}\biggr)^{1/4} \biggl(\frac{{\mathcal A}_{{\mathcal R}}}{2.41 \times 10^{-9}} \biggr)^{1/4} \biggl(\frac{h_{0}^2 \Omega_{R0}}{4.15 \times 10^{-5}}\biggr)^{1/4}.
\label{inf6}
\end{equation}
In Eq. (\ref{inf6})  $\gamma$ accounts for the possibility of a delayed reheating terminating at an Hubble scale $H_{r}$
smaller than the Hubble rate during inflation.

In what follows, as already mentioned in the introduction, we shall 
rather stick to the conventional case where the reheating is sudden and $\gamma = 1/2$ (or $H_{1} = H_{r}$ since the end of the 
inflationary phase coincides with the beginning of the radiation epoch).
In more general terms, however, 
 $H_{r}$ can be as low as $10^{-44} M_{P}$ (but not smaller) corresponding to a reheating scale occurring just prior to the formation of the light nuclei.  
 
In the limit $\alpha \to 0$ and $n_{1}\to 1$, Eq. (\ref{inf5}) leads to the standard result, namely:
\begin{equation}
\lim_{\alpha \to 0,\,\,\, n_{1}\to 1} {\mathcal P}_{T}(k,\tau) \to \frac{16}{\pi} \biggl(\frac{H_{1}}{M_{P}} \biggr)^2 \biggl(\frac{k}{k_{max}}\biggr)^{- 2 \epsilon},
\label{inf7}
\end{equation}
which implies $H_{1}/M_{P} = \pi \,r_{T} \,{\mathcal A}_{{\mathcal R}}/16$ with 
$r_{T} = 16 \epsilon$ and $n_{T} = - r_{T}/8$.  This kind of consistency relation (stipulating that the tensor scalar ratio exactly equals $16 \epsilon$) will not be valid anymore in the present context and the 
specific form of the tensor to scalar ratio will be used in section \ref{sec4} to constrain the possible values of $\alpha$. 

We observe that whenever $\alpha> 0$ the spectral index can increase since it can be naively larger than the slow-roll parameter. 
More specifically expanding $n_{T}(\epsilon,\alpha)$ in the limit $\epsilon <1$ we will have that 
\begin{equation}
n_{T}  = \frac{3\alpha}{1 + \alpha} + \frac{(\alpha -2)}{(1+ \alpha)^2} \epsilon + {\mathcal O}(\epsilon^2).
\label{inf8}
\end{equation}
Two possible situations can be envisaged. If $\alpha > 1$ the spectral slope in always violet (i.e. sharply increasing); in the limiting case $\alpha \gg 1$ we have, according to Eq. (\ref{inf8}), that $n_{T} \to 3$. This possibility is strongly constrained by backreaction effects as we shall specifically see in section \ref{sec4}.
If $0< \alpha < 1$ the spectra are blue (i.e. slightly increasing) provided $\alpha > 2\epsilon/( 3 + 5 \epsilon)$; in the opposite case (i.e.  $\alpha < 2\epsilon/( 3 + 5 \epsilon)$) the conventional limit is recovered and $n_{T} \simeq - 2 \epsilon$.

The physical region of the parameters 
corresponds to the situation where at the onset of inflation the refractive index is larger 
than (or equal to ) $1$. If this is the case a superluminal phase velocity 
is avoided.  Indeed, denoting with $\tau_{i}$ the initial time of the evolution we shall have that\footnote{We recall that the parametrization of the scale factors given in Eqs. (\ref{sfinf})--(\ref{sfmat}) stipulates that $a_{1} = a(- \tau_{1}) =1$. There are some who prefer 
to set $a_{0}=1$ (where $a_{0}$ is the present value of the scale factor) but this is not the convention adopted in the 
present paper}.  
\begin{equation}
n_{i} = n_{1} \biggl(\frac{a_{i}}{a_{1}}\biggr)^ {\alpha} = n_{1} e^{ - \alpha N_{t}}, \qquad n_{i} \geq 1.
\label{inf9}
\end{equation}
If $n_{i} =1$ we shall have that $n_{1} \to \exp{(\alpha N_{t})}$ where $N_{t}$ is the total 
number of inflationary efolds. As we shall see the detailed discussion of section \ref{sec5} implies that $n_{i}$ must indeed be ${\mathcal O}(1)$ even if not strictly equal to $1$. 
\subsection{Power spectrum during the radiation epoch}
Following the conventions established in Eqs. (\ref{inf1}) and (\ref{inf2}) the expression of the mode function during the radiation 
epoch shall be written as:
\begin{equation}
f_{k}(\tau) = \frac{{\mathcal D}_{r}}{\sqrt{2 \omega_{r}(\tau)}} \sqrt{\omega_{r}(\tau) \,y(\tau,\tau_1)}\,  \biggl\{ c_{+}(k, \tau_{1}) \, 
H_{\rho}^{(2)}[g_{r}(\tau)]+ c_{-}(k, \tau_{1})  H_{\rho}^{(1)}[g_{r}(\tau)]\biggr\},
\label{rad1}
\end{equation}
where 
\begin{equation}
 y(\tau,\tau_1) = \biggl(\tau + \frac{(\beta + 1)}{\beta} \tau_{1}\biggr), \qquad \rho = \frac{1}{2\, |1 - \alpha|}, \qquad 
 |{\mathcal D}_{r}|  = \sqrt{\frac{\pi}{2 |1 -\alpha|}}.
 \label{rad1a}
 \end{equation}
In terms of  $y(\tau,\tau_1)$ the argument of the Hankel functions $g_{r}(\tau)$ is defined as 
\begin{equation}
g_{r}(\tau) = \frac{\omega_{r}(\tau)}{| 1 - \alpha |} y(\tau,\tau_{1})  = \frac{k}{n_{1} | 1 -\alpha|} \biggl(\frac{\tau_{1}}{\beta}\biggr)^{\alpha} 
\biggl[\tau + \frac{(\beta + 1)}{\beta} \tau_{1}\biggr]^{1-\alpha}.
\label{rad2}
\end{equation}
The coefficients $c_{\pm}(k, \tau_{1})$ are complex and they obey $|c_{+}(k,\tau_{1})|^2 - |c_{-}(k,\tau_{1})|^2=1$. The exact expression 
of the two mixing coefficients is reported in Eqs. (\ref{cp1}) and (\ref{cm1}) of appendix \ref{APPA} and can be 
determined by matching continuously the inflationary mode function (i.e.  Eq. (\ref{inf1})) with the one of Eq. (\ref{rad1}) in $ \tau = - \tau_{1}$;
in formulae the following pair of conditions must be imposed:
\begin{equation}
f_{k}^{(\mathrm{inf})}( - \tau_{1}) = f_{k}^{(\mathrm{rad})}(-\tau_{1}), \qquad \frac{\partial f_{k}^{(\mathrm{inf})}}{\partial \tau} \biggl|_{\tau = - \tau_{1}} = 
\frac{\partial f_{k}^{(\mathrm{rad})}}{\partial \tau} \biggl|_{\tau = - \tau_{1}}.
\label{rad3}
\end{equation}
The requirements of Eq. (\ref{rad3}) follow directly from the continuity of the scale factors and of the extrinsic curvature. We remind 
that the continuity of the scale factor guarantees, in the present approach, the continuity of the refractive index. The 
continuity of the extrinsic curvature (related to the conformal time derivative of the scale factor) guarantees that $a^{\prime\prime}/a = 
{\mathcal H}^2 + {\mathcal H}^{\prime}$ will have, at most, jump discontinuities. The exact expression of the mixing 
coefficients determined in the present situation reproduces the conventional results when $\alpha\to 0$ (see Eq. (\ref{cpm}) 
of appendix \ref{APPA}).
 
The exact results of Eqs. (\ref{cp2}) and (\ref{cm2})  can be expanded in powers of $\overline{g}_{r}\simeq \overline{g}_{i} \ll 1$:
\begin{eqnarray}
c_{+}(k,\tau_{1}) - c_{-}(k,\tau_{1}) &=& {\mathcal O}( \overline{g}_{r}) + {\mathcal O}( \overline{g}_{i}),
\nonumber\\
c_{+}(k,\tau_{1}) + c_{-}(k,\tau_{1}) &=& 2 \,c_{-}(k,\tau_{1}) = \frac{2}{\overline{g}_{i}^{\mu} \,\overline{g}_{r}^{\rho}}\biggl[ 
i \, {\mathcal E}(\alpha, \epsilon) + {\mathcal O}( \overline{g}_{r}) + {\mathcal O}( \overline{g}_{i})\biggr],
\label{rad4}
\end{eqnarray}
where $\overline{g}_{r} = g_{r}(-\tau_{1})$ and $ \overline{g}_{i} = g_{i}(-\tau_{1})$ (see also Eq. (\ref{gg1}) of appendix \ref{APPA}). For practical reasons, the following combination
\begin{equation}
{\mathcal E}(\alpha, \epsilon) = \frac{ 2^{\mu + \rho} \Gamma(\mu) \Gamma(\rho) }{8 \sqrt{\beta} ( 1 - \alpha) \pi } \,\,\sqrt{\frac{|1 - \alpha|}{| 1 + \alpha \beta|} }
\,\,\{\beta [ 1 - 2 (1-\alpha) \rho - 2 \mu \alpha] + 1 - 2\mu\},
\nonumber
\end{equation}
has been introduced in Eq. (\ref{rad4}); note that ${\mathcal E}(\alpha, \epsilon) $ only depends on $\alpha$ and $\epsilon$ 
since all the other auxiliary variables (i.e. $\rho$ and $\mu$) are independent functions of $\alpha$ and $\epsilon$. 
The result of Eq. (\ref{rad4}) can be made more explicit by using the expressions of $\overline{g}_{i}$ and $\overline{g}_{r}$; to lowest order in $x_{1} = k\tau_{1}$, the approximate expression of $|c_{-}(x_{1})|^2$ is given by
\begin{equation}
|c_{-}(x_{1})|^2 = {\mathcal E}^{2}(\alpha,\epsilon) \, \beta^{2 \rho} |1 - \alpha|^{2\rho} \, | 1 + \alpha \beta|^{2 \mu} n_{1}^{- 2 (\mu + \rho)} \, x_{1}^{- 2 (\mu +\rho)}.
 \label{rad5a}
\end{equation} 
Since Eq. (\ref{rad4}) we have that $c_{+}(k,\tau_{1}) \simeq c_{-}(k,\tau_{1})$, for $\tau > - \tau_{1}$ the radiation power spectrum is therefore given by:
\begin{equation}
{\mathcal P}_{T}(k,\tau) = \frac{4k^3 \ell_{P}^2}{\pi | 1 -\alpha| a^2(\tau)} \,y(\tau,\tau_{1})\, |c_{-}(k, \tau_{1})|^2 \, J^2_{\rho}[g_{r}(\tau)],
\label{rad6}
\end{equation}
where $J_{\rho}(g_{r})= [H^{(1)}_{\rho}(g_{r}) + H^{(2)}_{\rho}(g_{r})]/2$.  The argument of $J_{\rho}[g_{r}(\tau)]$ is $g_{r}$ (not $\overline{g}_{r}$)
so that deep in the radiation epoch the power spectrum can be obtained in the limit $g_{r} \gg 1$. In this case, using the 
standard limits of the Bessel functions, the power spectrum becomes:
\begin{eqnarray}
{\mathcal P}_{T}(k,\tau) = \frac{64}{\pi} \biggl(\frac{H_{1}}{M_{P}}\biggr)^2 \biggl(\frac{a_{1}}{a}\biggr)^2 n_{r}(\tau)\, |k\tau_{1}|^2 \, |c_{-}(k,\tau_{1})|^2 \cos^2{[g_{r}(\tau)]};
\label{rad7}
\end{eqnarray}
where the large argument limit of $J_{\rho}[g_{r}(\tau)]$ has been used.
In Eq. (\ref{rad7}) we can replace $\cos^2{[g_{r}(\tau)]} \to 1/2$  as it is customary in this kind of analyses.
Thus, from Eq. (\ref{rad7})  we can also deduce the energy density and express it in critical units:
\begin{equation}
\Omega_{gw}(k,\tau) = \frac{8}{3 \pi} \biggl(\frac{H_{1}}{M_{P}}\biggr)^2  \frac{ |k\tau_{1}|^4}{n_{r}(\tau)} |c_{-}(k,\tau_{1})|^2.
\label{rad9}
\end{equation}
\subsection{Power spectrum during the matter epoch} 
In the matter-dominated epoch, using Eq. (\ref{sfmat}) into Eq. (\ref{opt1}), the normalized solution of Eq.  (\ref{MODEB}) for $\tau \geq \tau_{2}$  is given by:
 \begin{equation}
f_{k}(\tau) = \frac{{\mathcal D}_{m}}{ \sqrt{2 \omega_{m}(\tau)}} \sqrt{ \omega_{m}(\tau)\, z(\tau,\tau_{1},\tau_{2})} \biggl\{d_{+}(k,\tau_{1},\tau_{2}) \, 
H_{\sigma}^{(2)}[g_{m}(\tau)] + d_{-}(k,\tau_{1},\tau_{2}) H_{\sigma}^{(1)}[g_{m}(\tau)] \biggr\},
\label{mat1}
\end{equation}
where
\begin{equation}
z(\tau,\tau_{1}, \tau_{2}) =  \biggl(\tau + \tau_{2} + \frac{2 (\beta + 1)}{\beta} \tau_{1}\biggr), \qquad  \sigma = \frac{3}{2\, |1 - 2 \alpha|}, 
\qquad |{\mathcal D}_{m}|  = \sqrt{\frac{\pi}{2 |1 -2\alpha|}}.
 \label{mat2}
\end{equation} 
Notice finally that $g_{m}(\tau)$ in Eq. (\ref{mat1}) is defined as
\begin{eqnarray}
g_{m}(\tau) &=& \frac{\omega_{m}(\tau)}{| 1 - 2 \alpha |} \biggl(\tau + \tau_{2} +  \frac{2(\beta + 1)}{\beta} \tau_{1}\biggr)
\nonumber\\
 &=& \frac{k}{n_{1} | 1 -2\alpha|} 
\biggl\{ \frac{4 \tau_{1} [\beta\tau_{2} + (\beta + 1)\tau_{1}]}{\beta^2} \biggr\}^{\alpha} 
\biggl[ \tau + \tau_{2} + \frac{2(\beta + 1)}{\beta} \tau_{1} \biggr]^{ 1 - 2\alpha}.
\label{mat3}
\end{eqnarray}
As in the case of $c_{\pm}(k,\tau_{1})$ also the mixing coefficients $d_{\pm}(k, \tau_{1}, \tau_{2})$ can be determined by continuous
matching of the relevant mode functions across $\tau_{2}$. In this case we shall then impose 
\begin{equation}
f_{k}^{(\mathrm{rad})}(\tau_{2}) = f_{k}^{(\mathrm{mat})}(\tau_{2}), \qquad \frac{\partial f_{k}^{(\mathrm{rad})}}{\partial \tau} \biggl|_{\tau =  \tau_{2}} = 
\frac{\partial f_{k}^{(\mathrm{mat})}}{\partial \tau} \biggl|_{\tau = \tau_{2}}.
\label{mat4}
\end{equation}
The explicit form of $d_{\pm}(k,\tau_{1},\tau_{2})$ is reported 
in Eqs. (\ref{dp1}) and (\ref{dm1}) of appendix \ref{APPB} together with a specific discussion of some relevant physical limits.

As in the case of radiation the power spectrum can be easily obtained in all the interesting regions.
More specifically, recalling Eqs. (\ref{dp1}) and (\ref{dm1}) we have that $d_{+}(k,\tau_{1}, \tau_{2}) \simeq d_{-}(k,\tau_{1}, \tau_{2})$
in the relevant physical limit. From Eq. (\ref{rad3}) and (\ref{mat4})  the power spectrum during the matter phase is therefore given by:
\begin{equation}
{\mathcal P}_{T}(k,\tau) = \frac{4k^3 \ell_{P}^2}{\pi | 1 -2 \alpha| a^2(\tau)} \,z(\tau,\tau_{1}, \tau_{2})\, |d_{-}(k, \tau_{1}, \tau_{2})|^2 \, J^2_{\sigma}[g_{m}(\tau)],
\label{mat5}
\end{equation}
where $J_{\sigma}(g_{m})= [H^{(1)}_{\rho}(g_{m}) + H^{(2)}_{\rho}(g_{m})]/2$.  Deep in the matter epoch the power spectrum can be obtained in the limit $g_{m} \gg 1$. In this case, using the 
standard limits of the Bessel functions, the power spectrum becomes:
 \begin{eqnarray}
{\mathcal P}_{T}(k,\tau) = \frac{64}{\pi} \biggl(\frac{H_{1}}{M_{P}}\biggr)^2\, \biggl(\frac{a_{1}}{a}\biggr)^2 \, n_{m}(\tau)\, |k\tau_{1}|^2 \, |d_{-}(k,\tau_{1},\tau_{2})|^2\, \cos^2{[g_{m}(\tau)]}.
\label{mat6}
\end{eqnarray}
In Eq. (\ref{mat6}) we can replace $\cos^2{[g_{m}(\tau)]} \to 1/2$  and  we can also deduce the energy density and express it in critical units:
\begin{equation}
\Omega_{gw}(k,\tau) = \frac{8}{3 \pi} \biggl(\frac{H_{1}}{M_{P}}\biggr)^2  \,  \biggl(\frac{H_{0}}{H_{eq}}\biggr)^{2/3} \,\frac{ |k\tau_{1}|^4}{n_{m}(\tau)} |d_{-}(k,\tau_{1},\tau_{2})|^2.
\label{mat7}
\end{equation}
Using the same expansions discussed in the radiation case 
we can obtain, for instance, the leading order expression for $|d_{-}(k, \tau_{1}, \tau_{2})|^2$ for $|k \tau_{1}|\ll 1$ and $|k \tau_{2}|\ll 1$:
\begin{equation}
|d_{-}(k, \tau_{1}, \tau_{2})|^2 = {\mathcal M}(\alpha,\epsilon) \biggl(\frac{\tau_{1}}{\beta \tau_{2}}\biggr)^{ 2 \alpha(\rho - \sigma)} 
n_{1}^{ 2 (\mu + \sigma)} \, |k\tau_{1}|^{- 2 (\mu + \rho)} \, |k\tau_{2}|^{ 2 (\rho - \sigma)},
\label{mat8}
\end{equation}
where ${\mathcal M}(\alpha,\epsilon)$ is given by:
\begin{eqnarray}
{\mathcal M}(\alpha,\epsilon) &=& \frac{2^{2 \mu - 9} \beta^{1 + \rho} \Gamma^2(\sigma) \Gamma^2(\mu)[1 + \sigma ( 1 - 2 \alpha) + 2 \rho ( 1 - \alpha)]^2}{\pi^2 \rho^2 |1 - \alpha|^2 \,|1 - 2 \alpha|^{1 - 2 \sigma}\, |1 + \alpha \beta|^{ 1 - 2\mu} }
\nonumber\\
 &\times& \biggl[ \frac{\beta+ 1}{2 \beta} - ( 1 - \alpha) \rho - \mu \frac{(1 + \alpha \beta)}{\beta} \biggr]^2,
\end{eqnarray}
and it is only function of $\alpha$ and $\epsilon$ because the other parameters (i.e. $\mu$, $\rho$ and $\sigma$) are all independent functions of $\alpha$ and $\epsilon$ and they have been defined, respectively, in Eqs. (\ref{inf1}), (\ref{rad1a}) and (\ref{mat2}). 

An interesting limit of Eq. (\ref{mat8}) is the one $\beta \to 1$ and $\alpha \to 0$; in this limit $\mu \to 3/2$, $\rho \to 1/2$ and $\sigma \to 3/2$. In this case 
(setting also $n_{1} \to 1$) we have that $|d_{-}|^2 \to (9/64) |k \tau_{1}|^{-4} |k \tau_{2}|^{-2}$. This is the standard result for the mixing coefficients in the 
case of a transition from a pure de Sitter phase to the matter-dominated epoch passing through the conventional radiation dominance\footnote{The limit of the exact expressions in this specific case (see Eqs. (\ref{dpex}) and (\ref{dmex})) coincides with the limit of the general expression obtained above: this is a useful check of the whole algebraic 
consistency.}. 

It is finally possible to obtain a general expression encompassing the radiation and matter-dominated phases 
for the energy density of the relic gravitons in critical units. The expressions applies for modes inside the Hubble 
radius at the present time and it is given by:
\begin{eqnarray}
\Omega_{gw}(k,\tau_0) &=& {\mathcal N}(\alpha, \epsilon) \biggl(\frac{H_{0}}{H_{eq}}\biggr)^{2/3+ \alpha [1/6 - (\rho -\sigma)]}
\biggl(\frac{H_{1}}{M_{P}}\biggr)^{ 2 - \alpha [ 2 (\rho - \sigma) +1]/2} \biggl(\frac{H_{0}}{M_{P}}\biggr)^{\alpha[2 (\rho - \sigma) +1]/2}
\nonumber\\
&\times& n_{i}^{2 ( \mu + \sigma) -\alpha} e^{ N_{t} \alpha [ 2 (\mu +\sigma) - \alpha]} 
\biggl(\frac{k}{k_{max}}\biggr)^{4 - 2 (\mu + \rho)}\, 
{\mathcal T}(k,k_{eq}), 
\label{OMFIN}\\
{\mathcal T}^2(k,k_{eq}) &=& 1 + b_{1} \biggl(\frac{k}{k_{eq}}\biggr)^{ 2(\rho - \sigma)} + c_{1} \biggl(\frac{k}{k_{eq}}\biggr)^{4 (\rho - \sigma)},
\nonumber\\
{\mathcal N}(\alpha, \epsilon) &=&  \frac{2^{2 \alpha+ 3} }{3\pi} \beta^{- \alpha [1 + 2 (\rho -\sigma)]} \,{\mathcal M}(\alpha,\epsilon),
\nonumber
 \end{eqnarray}
 where $b_{1}$ and $c_{1}$ are numerical constants of ${\mathcal O}(1)$; in Eq. (\ref{OMFIN}) we have already expressed $n_{1}$ in terms of $n_{i}$, i.e. the initial value of the spectral index. The accurate 
 value of $b_{1}$ and $c_{1}$ can be also obtained numerically by computing the transfer 
 function of the energy density of the relic gravitons introduced in (see e.g. \cite{max12}). Alternatively one can compute the transfer function for the 
 power spectrum and then compute the energy density \cite{FS1}. In both cases 
 the idea is to integrate numerically the background and the mode functions across the matter-radiation transition. For the 
 present ends what matters, however, is Eq. (\ref{OMFIN}) in the limit $ k \gg k_{eq}$. 
 When $\alpha \to 0$ and $n_{i}\to 1$ and for 
 $ k \gg k_{eq}$ Eq. (\ref{OMFIN}) reproduces the standard result
 \begin{equation}
 \Omega_{gw}(k,\tau_0) =\frac{3}{8 \pi}  \biggl(\frac{H_{0}}{H_{eq}}\biggr)^{2/3}
\biggl(\frac{H_{1}}{M_{P}}\biggr)^{ 2} \biggl(\frac{k}{k_{max}}\biggr)^{ - 2 \epsilon}.  
\label{stan}
\end{equation}
As in the standard case, when $k \gg k_{max}$ we have that $\Omega_{gw}$ is exponentially suppressed as $\exp{[- \delta k/k_{max}]}$ 
where $\delta$ is a numerical factor that can be estimated in a specific model of smooth transition (see. e.g. \cite{max12}).

In Eq. (\ref{OMFIN}) the ratio
$H_{0}/M_{P}$ is raised to an $\alpha$-dependent power that disappears 
in the conventional limit of Eq. (\ref{stan}) (i.e. $\alpha \to 0$). 
The explicit values of $H_{0}/M_{P}$ and $H_{eq}/H_{0}$ can be explicitly written as 
 \begin{eqnarray}
 \frac{H_{0}}{M_{P}} &=& 1.228 \times 10^{-61} \biggl(\frac{h_{0}}{0.7}\biggr),\qquad \frac{H_{\mathrm{eq}}}{H_{0}} = 10^{5.27}  \biggl(\frac{h_{0}^2 \Omega_{M0}}{0.1364}\biggr)^{3/2}  \biggl(\frac{h_{0}^2 \Omega_{R 0}}{4.15 \times 10^{-5}}\biggr)^{-3/2},
 \label{num}
 \end{eqnarray}
where $h_{0}$ is the indetermination on the present value of the Hubble rate, $\Omega_{M0}$ is the critical fraction 
of matter density and $\Omega_{R0}$ is the critical fraction of radiation energy density. 

It is finally useful to remark that the spectral slope of $\Omega_{gw}$ in the high-frequency branch (i.e. for $k \gg k_{eq}$) 
is simply given by $ 4 - 2 (\mu+ \rho)$ as it can be immediately verified from the explicit expression of Eq. (\ref{OMFIN}).
Recalling  Eqs. (\ref{inf1}) and (\ref{rad1}) the high-frequency slope can be written more explicitly:
\begin{equation}
 4 - 2 [\mu(\epsilon,\alpha)+ \rho(\alpha)]  = 3 + \frac{1}{\alpha -1} + \frac{\alpha - 2}{1 + \alpha - \epsilon} \simeq 2 (\alpha - \epsilon) + {\mathcal O}(\epsilon^2) + {\mathcal O}(\alpha^2), 
\label{SPom} 
 \end{equation}
where the second equality follows by expanding the exact expression first in powers of $\epsilon$ and then in powers of $\alpha$.
This limit captures an important corner of the parameter 
space (see the discussion of section \ref{sec4}).  Equation (\ref{SPom}) implies that in the pure de Sitter limit without variation of the spectral index (i.e. $\alpha =0$ and $\epsilon =0$) $\Omega_{gw}$ is constant in frequency with amplitude given by Eq. (\ref{stan}) in the limit $k \gg k_{eq}$.

\subsection{Typical frequencies}
We shall always use wavenumbers\footnote{We shall often measure comoving wavenumbers in Hz and refer to typical comoving frequencies. Note that, in natural units, $k = 2\pi \nu$. Frequencies and wavenumbers are not exactly coincident even if it is useful, at a practical level, to measure wavenumbers in Hz.} expressed either in units of $\mathrm{Mpc}^{-1}$ or in units of Hz. The reason 
of this potential ambiguity is that the discussion mixes constraints arising over large length-scales (where the wavenumber are typically 
measured in $\mathrm{Mpc}^{-1}$) and other limits coming from comparatively much shorter scales (where the wavenumbers 
ate typically assigned in Hz) \cite{max12}.  For instance, in what follows
we shall be dealing with the big-bang nucleosynthesis wavenumber $k_{bbn}$ 
\begin{equation}
k_{bbn}= 
1.47\times 10^{-10} \biggl(\frac{g_{\rho}}{10.75}\biggr)^{1/4} \biggl(\frac{T_{bbn}}{\,\,\mathrm{MeV}}\biggr) 
\biggl(\frac{h_{0}^2 \Omega_{\mathrm{R}0}}{4.15 \times 10^{-5}}\biggr)^{1/4}\,\,\mathrm{Hz},
\label{kbbn}
\end{equation}
where $g_{\rho}$ denotes the effective number of relativistic degrees of freedom entering the total energy density of the plasma and $T_{bbn}$ is the 
big-bang nucleosythesis temperature determining the size of the Hubble radius at the corresponding epoch. 
The typical value of the frequency corresponding to Eq. (ref{kbbn}) is $\nu_{bbn} = k_{bbn}/2\pi= 2.3\times 10^{-11}$. Similar observations can be made in all the other cases.  For future convenience $k_{max}$ and $k_{eq}$ can also be expressed in Hz:
\begin{eqnarray}
k_{max} &=& 2.183 \biggl(\frac{H_{r}}{H}\biggr)^{\gamma -1/2} \biggl(\frac{\epsilon}{0.01}\biggr)^{1/4} \biggl(\frac{{\mathcal A}_{{\mathcal R}}}{2.41 \times 10^{-9}} \biggr)^{1/4} \biggl(\frac{h_{0}^2 \Omega_{R 0}}{4.15 \times 10^{-5}}\biggr)^{1/4}\,\,\, \mathrm{GHz},
\label{kmax}\\
k_{eq} &=& 9.69 \times 10^{-17} \, \biggl(\frac{h_{0}^2 \Omega_{M0}}{0.1364}\biggr)\, \biggl(\frac{h_{0}^2 \Omega_{R 0}}{4.15 \times 10^{-5}}\biggr)^{-1/2}\,\,\,\mathrm{Hz}.
 \label{keq}
 \end{eqnarray}
The frequencies corresponding to the fiducial values of the parameters given in Eqs. (\ref{kmax}) and (\ref{keq}) are given, respectively, by 
$\nu_{max} = 0.34\, \mathrm{GHz}$ and by $\nu_{eq} =1.54\times 10^{-17}$ Hz.
 \subsection{Secondary effects}
In the present analysis we neglected, for the sake of simplicity, a number of secondary effects 
that may interfere with the variation of the refractive index. 
For $k< k_{bbn}$ the power spectra and the energy density of the gravitons are suppressed due to the neutrino free 
streaming. The effective energy-momentum tensor acquires, to first-order in the amplitude 
of the plasma fluctuations, an anisotropic stress (see e. g. \cite{wnu1} and references therein). 
The overall effect of collisionless particles is a reduction 
of the spectral energy density of the relic gravitons\footnote{ Assuming that the only collisionless 
species in the thermal history of the Universe are the neutrinos, the amount 
of suppression can be parametrized by the function ${\mathcal F}(R_{\nu}) = 1 -0.539 R_{\nu} + 0.134 R_{\nu}^2$,
where $R_{\nu}$ is the fraction of neutrinos in the radiation plasma.  In the case $N_{\nu} = 3$, 
$R_{\nu} = 0.405$ and the suppression of the spectral energy density is proportional 
to ${\mathcal F}^2(0.405)= 0. 645$. This suppression will be effective for relatively 
small frequencies which are larger than $k_{eq}$ and smaller than $k_{bbn}$.}.

The second effect leading to a further suppression of the energy density is the late dominance of the dark energy. The redshift of $\Lambda$-dominance is given by $\Omega_{\mathrm{de}}/\Omega_{\mathrm{M}0}$.
In principle there should be a break in the spectrum for the modes reentering the Hubble radius after $\tau_{\Lambda}$.  This tiny  modification of the slope is practically irrelevant and it occurs anyway for $k < k_{eq}$. However,  the adiabatic damping of the tensor mode function across the $\tau_{\Lambda}$-boundary reduces the amplitude of the spectral energy density by a factor $(\Omega_{\mathrm{M}0}/\Omega_{\Lambda})^2 \simeq 0.10$. This figure is comparable with the suppression due to 
the neutrino free streaming. These effects have been discussed in the past (see \cite{TR1,max12} and references therein).  
Further effects leading to similar reductions of $\Omega_{gw}$ are related to the evolution of the 
relativistic species. 

The effects mentioned in the two previous paragraphs are secondary since they can be easily reabsorbed by the variation of one of the other unknown parameters of the cosmic graviton background. At the same time they become truly essential if the absolute normalization of the graviton spectrum is known (see last paper of Ref. \cite{max12}). 
In the present case the inclusion of these secondary effects is unimportant for the final conclusions, as we explicitly 
checked.

\renewcommand{\theequation}{4.\arabic{equation}}
\setcounter{equation}{0}
\section{Phenomenological considerations}
\label{sec4}
The limits on the variation of the refractive index over various scales will now be derived.
There are four qualitatively different sets of bounds to be examined and they involve, respectively, {\it (i)} the backreaction 
constraints during inflation, {\it (ii)} the limits stemming from the tensor to scalar ratio obtained from the temperature and polarization anisotropies of the cosmic microwave background, {\it (iii)} the bounds arising from the millisecond pulsar timing measurements and finally 
{\it (iv)} the so-called big-bang nucleosynthesis constraints. At the end of the section the impact of the derived limits on the prospects for the wide-band interferometers shall be addressed.  

\subsection{Limits from backreaction effects}
The considerations of section \ref{sec3} suggesting an upper limit on  $\alpha$ can be 
made more concrete by computing the total energy density of the gravitational waves and by comparing it with 
the critical energy density during inflation.  From Eqs. (\ref{T8e}) and (\ref{T12}) the total energy density of the produced gravitons in critical units  is given by:
\begin{equation}
\frac{\rho_{gw}(a,\alpha,\epsilon)}{\rho_{crit}}= \frac{1}{24 H_{1}^2 a^2}\int_{1/\tau_{i}}^{1/\tau_{1}} \frac{d k}{k} \, \biggl[ \frac{k^2}{n^2(\tau)} {\mathcal P}_{T}(k, \tau) + {\mathcal Q}_{T}(k,\tau)\biggr],
\qquad \rho_{crit} = \frac{3 H_{1}^2 M_{P}^2}{8\pi},
\label{endens1}
\end{equation}
where the integration is extended from modes the exiting the horizon at the onset of inflation up to those 
reentering exactly at the onset of the radiation phase. Inserting Eqs. (\ref{inf1}), (\ref{inf2}) and (\ref{inf3}) into Eq. (\ref{endens1}) 
and performing the indicated integrals we can easily obtain the following result\footnote{The cases $ \mu = 3/2$ and $
\mu = 5/2$ are singular: this simply means that the corresponding integrals must 
be separately computed and lead to a logarithmic contribution which is only present, strictly speaking in the case $\epsilon \to 0$ 
and $\alpha \to 0$ (i.e. pure de Sitter evolution). }
\begin{eqnarray}
\frac{\rho_{gw}(a,\, \alpha,\, \epsilon)}{\rho_{crit}} &=& {\mathcal S}(\alpha, \epsilon) \biggl(\frac{H_{1}}{M_{P}}\biggr)^2 \biggl(\frac{a}{a_{i}}\biggr)^{2 \mu \alpha}\biggl\{ \biggl(\frac{a_{1}}{a}\biggr)^{(3- 2\mu)/\beta}  \biggl[ 1 + \frac{(3 - 2 \mu)}{(5 - 2 \mu) \, n_{i}^2} \biggl(\frac{a}{a_{i}}\biggr)^{-2\alpha} \biggl(\frac{a_{1}}{a}\biggr)^{2/\beta} \biggr]
\nonumber\\
&-&  \biggl(\frac{a_{i}}{a}\biggr)^{(3- 2\mu)/\beta}\biggl[ 1 + \frac{(3 - 2 \mu)}{(5 - 2 \mu) \, n_{i}^2} \biggl(\frac{a}{a_{i}}\biggr)^{-2\alpha} \biggl(\frac{a_{i}}{a}\biggr)^{2/\beta} \biggr]\biggr\},
\nonumber\\
{\mathcal S}(\alpha, \epsilon) &=& \frac{2^{ 2 \mu} \, \Gamma^2(\mu) \, |1 + \alpha \beta|^{2 \mu -1} }{3 \,\pi^2 \,( 3 - 2 \mu) \,n_{i}^2}.
\label{endens2}
\end{eqnarray}
The function ${\mathcal S}(\alpha, \epsilon)$ appearing in Eq. (\ref{endens2}) only depends on $\alpha$ and $\epsilon$ since $\beta= \beta(\epsilon)$ and $ \mu = \mu(\alpha,\epsilon)$; moreover, the dependence on the scale factor in  Eq. (\ref{endens2}) can be traded for the total number of inflationary efolds $N_{t}$.

It is not necessary to analyze the independent variation of $N_{t}$, $\alpha$ and $\epsilon$:
 the upper bound on $\alpha$ is anyway less constraining than the 
ones to be examined later on. In fact $\alpha$ cannot exceed 
$0.1$ when the remaining parameters are fixed to their fiducial values: from Eq. (\ref{endens2}) with $a= a_{f} = a_{1}$ and $n_{i} = {\mathcal O}(1)$ we have that $|\rho_{gw}/\rho_{crit} |<1$ provided 
\begin{equation}
\alpha < - \frac{\ln{(\pi \, \epsilon \, {\mathcal A}_{{\mathcal R}})}}{ 3 \, N_{t}},  
\label{endens3}
\end{equation}
where, as usual,  ${\mathcal A}_{{\mathcal R}}$ is the amplitude of the scalar power spectrum at the pivot scale and has been introduced in Eq. (\ref{ST1Ca}). The total number of efolds $N_{t}$ appearing in Eq. (\ref{rT1}) must be larger than (or equal to) $N_{max}$:
\begin{equation}
 N_{max} = 61.43 + \frac{1}{4} \ln{\biggl(\frac{h_{0}^2 \Omega_{R 0}}{4.15 \times 10^{-5}} \biggr)} - \ln{\biggl(\frac{h_{0}}{0.7}\biggr)} + \frac{1}{4} \ln{\biggl(\frac{{\mathcal A}_{{\mathcal R}}}{2.41 \times 10^{-9}}\biggr)} + \frac{1}{4} \ln{\biggl(\frac{\epsilon}{0.01}\biggr)},
 \label{rT1a}
\end{equation} 
which is the maximal number of efolds presently accessible to large-scale observations\footnote{In practice $N_{max}$ is determined 
by redshifting the inflationary event horizon at the present time and by identifying the obtained results with the current value 
of the Hubble radius.}\cite{LL}. In the case $(\epsilon,\,N_{t},\, {\mathcal A}_{{\mathcal R}}) = (0.01,\, 65,\,  2.41\times 10^{-9})$ Eq. (\ref{rT1a}) implies, for instance,
$\alpha < 0.11$ when $n_{i} = {\mathcal O}(1)$. Similar results can be obtained from slightly different choices of parameters. 
\begin{figure}[!ht]
\centering
\includegraphics[height=8cm]{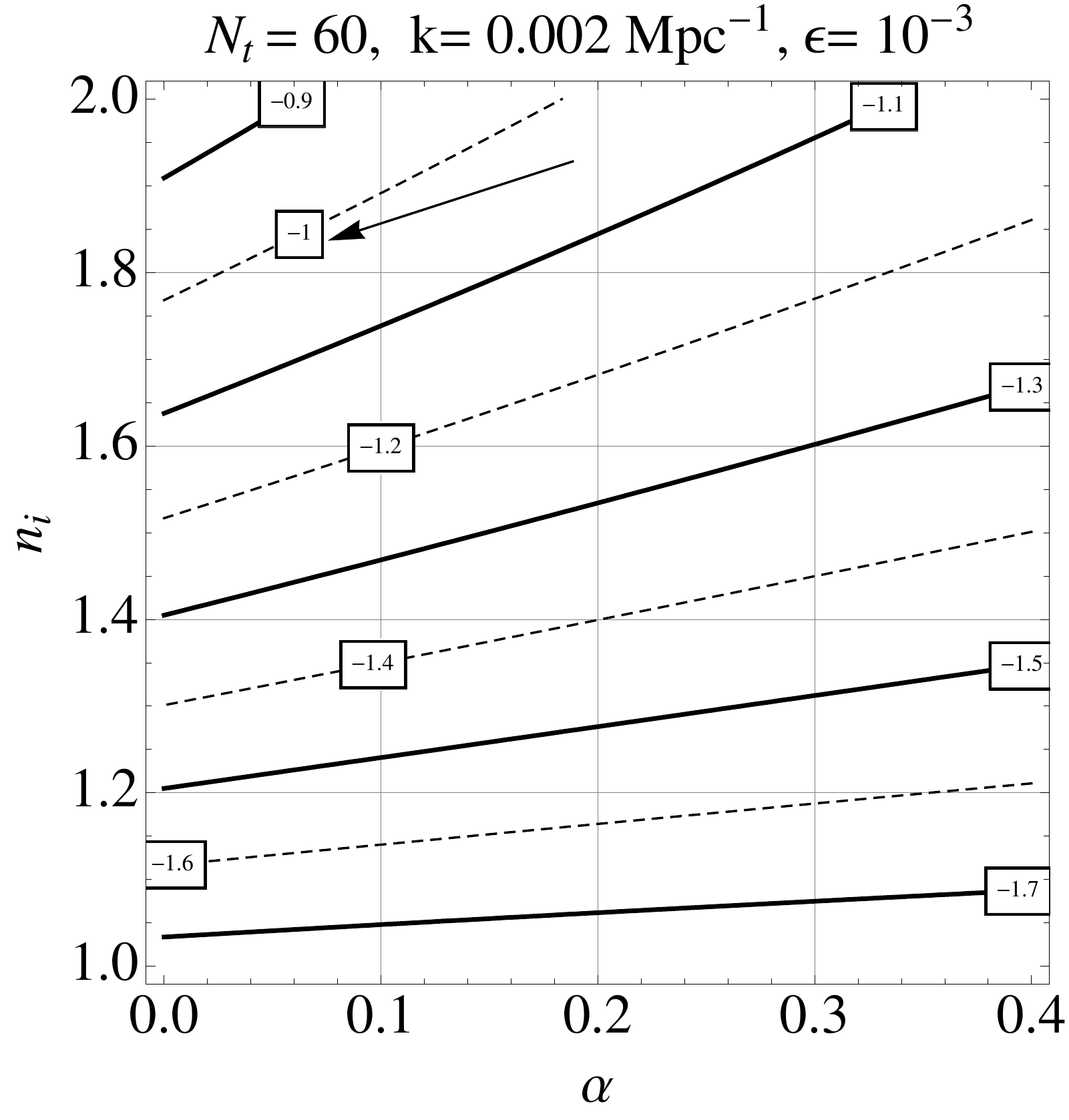}
\includegraphics[height=8cm]{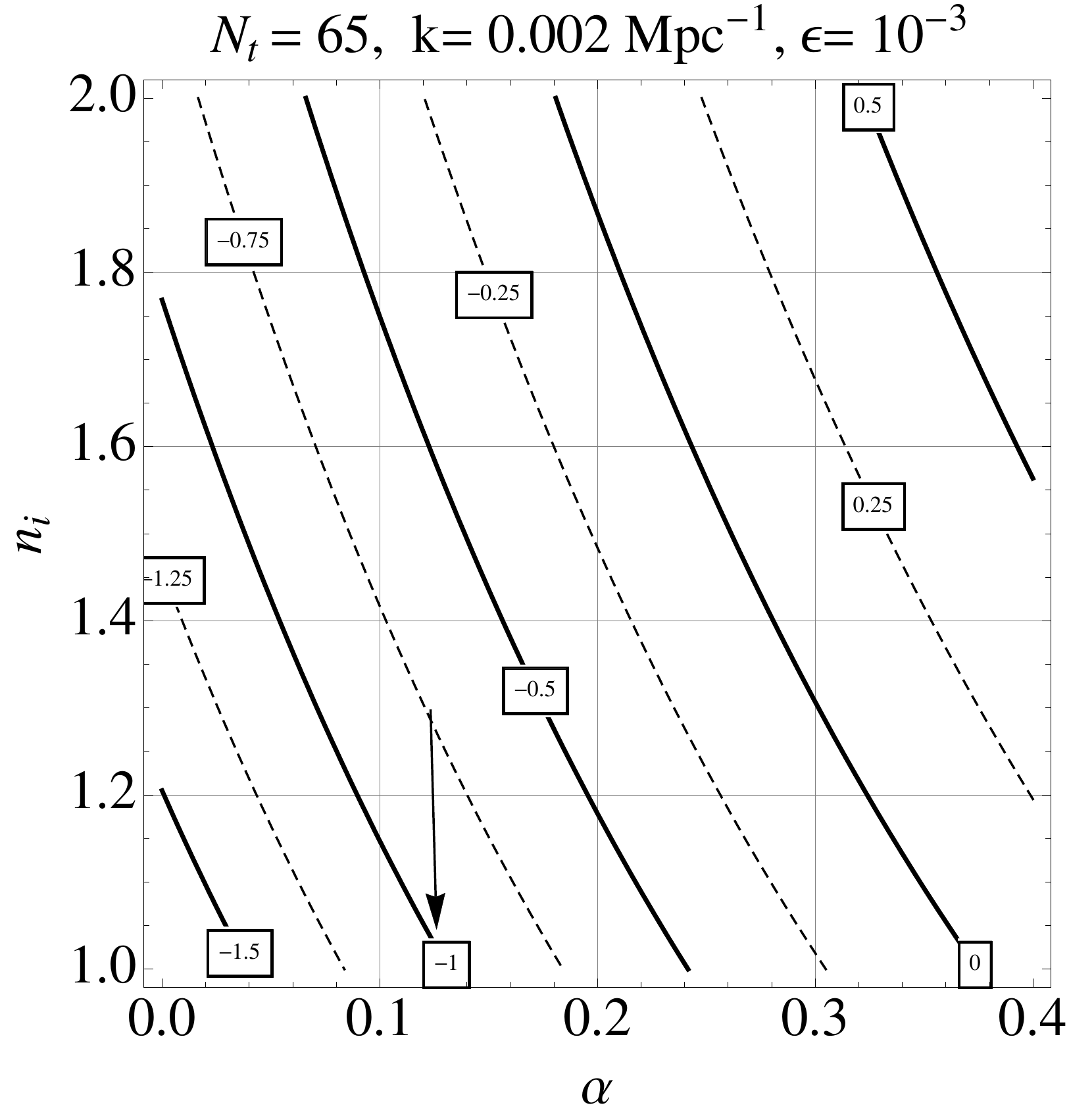}
\caption[a]{The value of $r_{T}$ computed from Eq. (\ref{rT1}) is illustrated in the $(\alpha, \, n_{i})$ plane.}
\label{Figure1}      
\end{figure}

\subsection{Limits from the tensor to scalar ratio}
The long wavelength gravitons induce direct temperature and polarization. Technically they can affect the $TT$ power spectra (i.e. the temperature autocorrelations) the $EE$ power spectra (i.e. the polarization autocorrelations) and the $TE$ power spectra (i.e. the cross-correlation between temperature and polarization). These power spectra  can interfere with temperature and polarization power spectra 
induced by the scalar mode and this is why the upper bounds on the tensor to scalar ratio can be derived from the accurate determinations of the temperature anisotropies and polarization anisotropies \cite{cmb1,cmb2}. Depending on the combined data sets the WMAP 5-year data provided bounds on $r_{T}$ in the framework of the concordance model with values ranging from $r_{T} < 0.58$ to $r_{T} < 0.2$. Similar bounds have been obtained 
from the WMAP 7-year data. The 9-year WMAP data release gave a limit  $r_{T} < 0.38$  always in the light of the concordance model in the presence of tensor. In the last three years there have been more direct determinations of the $B$-mode polarization of the cosmic microwave background. The first detection of a B-mode polarization, not caused by relic gravitons but coming from the lensing of the E-mode polarization, has been published  by the South Pole Telescope \cite{SPTpol}. The Bicep2 experiment \cite{bicep2} claimed the observation of a $B$-mode component with $r_{T}=0.2^{+0.07}_{-0.05}$ which turned out to be induced, at least predominatly, by a polarized foreground.  The present Planck data imply $r_{T} < 0.1$ \cite{cmb3} but the reported 
sensitivity to the $B$-mode polarization is rather poor.  

While tensor contribution 
to the cosmic microwave background observables can be cleanly ruled out (or ruled in) by direct observations of the $B$-mode 
polarization (as attempted by Bicep2 and by other previous experiments directly sensitive to polarization (see e.g. \cite{prev})) for the present ends what matters is not the specific value of the bound but the generic order of magnitude that  $r_{T}$ should not exceed at 
a conventional pivot scale\footnote{ The WMAP collaboration
consistently chooses $k_{p}= 0.002\, \, \mathrm{Mpc}^{-1}$. The Bicep2 collaboration used $k_{p} =0.05\, \mathrm{Mpc}^{-1}$.
The  first data release of the Planck collaboration assigned the scalar power spectra of curvature perturbations ${\mathcal P}_{{\mathcal R}}$ 
at $k_{p} =0.05\, \mathrm{Mpc}^{-1}$ while the tensor to scalar ratio $r_{T}$ is assigned at $k_{p} =0.002\, \mathrm{Mpc}^{-1}$.}
$k_{p} = 0.002\, \, \mathrm{Mpc}^{-1}$. From  Eq. (\ref{inf5})
the tensor to scalar ratio reads
\begin{equation}
r_{T}(k_{p}, \alpha, n_{i}, \epsilon, N_{t}) = \pi \,\,\epsilon\,\, {\mathcal C}(\epsilon,\alpha)\,\,n_{i}^{3 - n_T(\epsilon,\alpha)}\, \,e^{ 2 \alpha [3 - n_{T}(\epsilon,\alpha)] N_{t}}\,\, \biggl(\frac{k_{p}}{k_{max}}\biggr)^{n_{T}(\epsilon,\alpha)}.
\label{rT1}
\end{equation}
To get a superficial idea of the orders of magnitude involved we can first consider the case 
$N_{t} = 65$ and $\epsilon = {\mathcal O}(10^{-3})$. In this case we have that $r_{T} < 0.1$ provided $0< \alpha < 0.13$
for $n_{i} = 1$. A slight increase of $n_{i}$ or of the total number of efolds strengthen the limit on $\alpha$. For instance 
if $n_{i} =1.5$  we will have that $0<\alpha < 0.04$ (for  $N_{t} = 65$) and  $0<\alpha < 0.02$ (for $N_{t} = 70$). 
\begin{figure}[!ht]
\centering
\includegraphics[height=8cm]{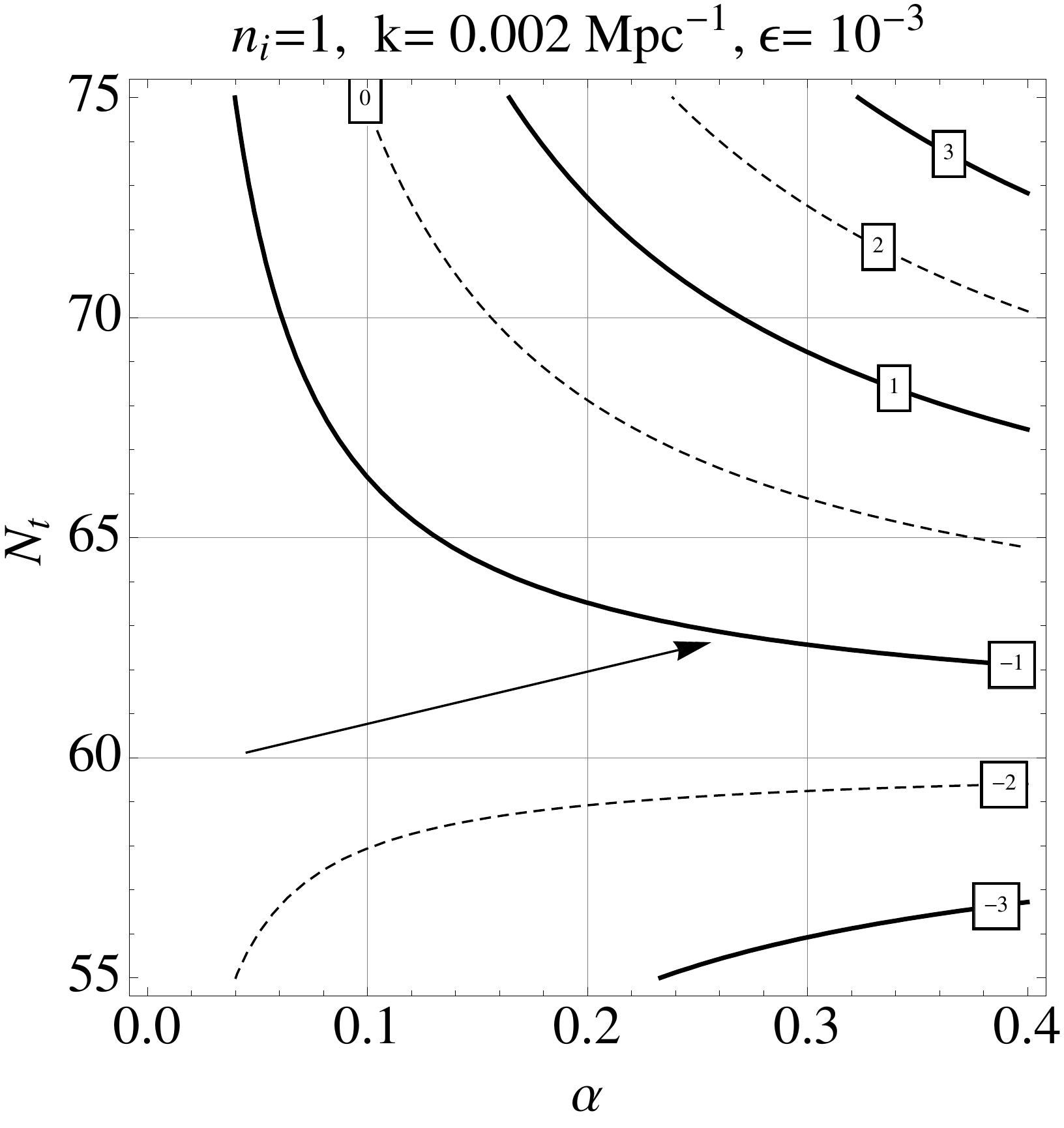}
\includegraphics[height=8cm]{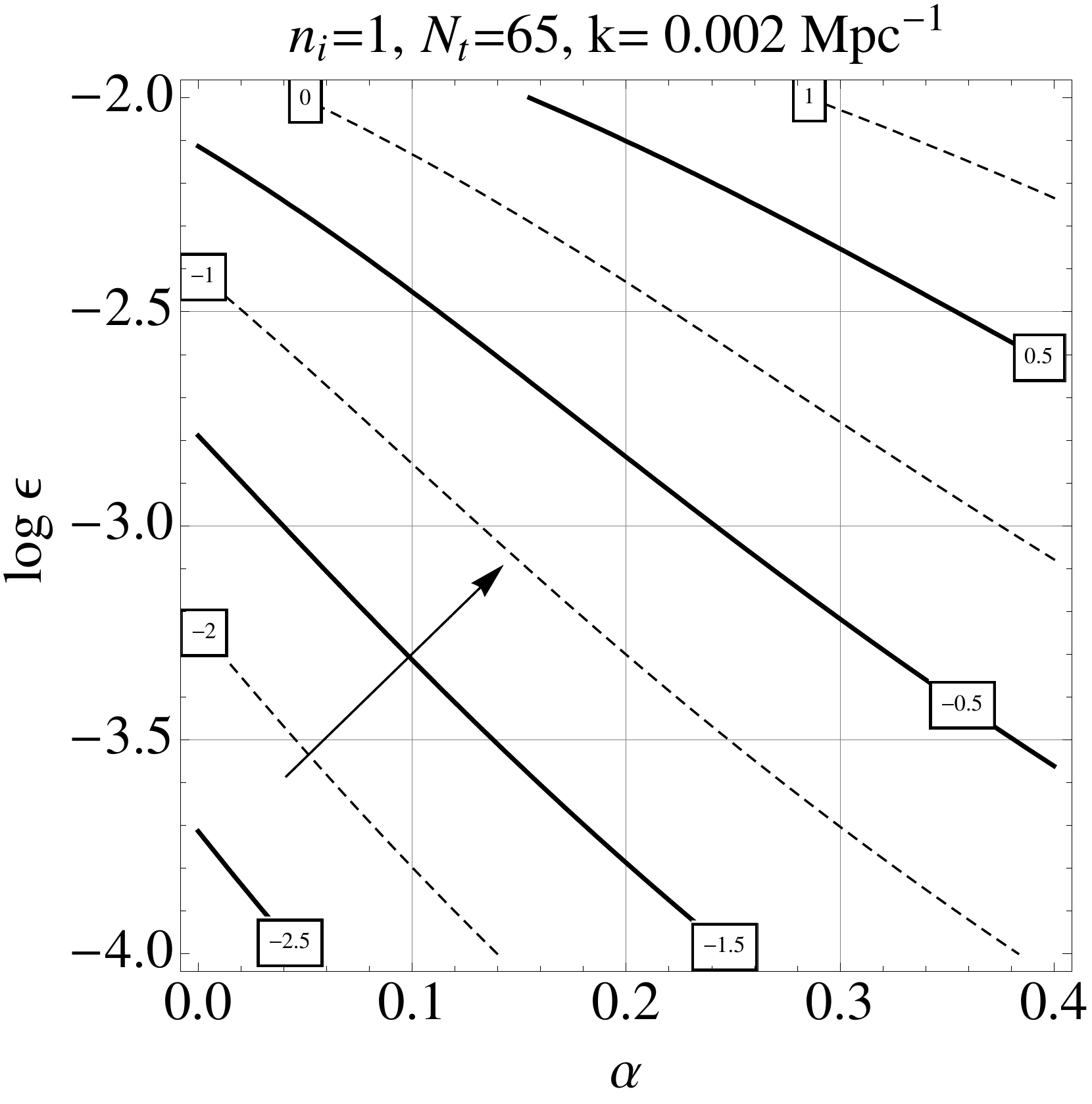}
\caption[a]{The value of $r_{T}$ is illustrated in the $(\alpha, \, N_{t})$ plane and in the $(\alpha, \log{\epsilon})$ plane.}
\label{Figure2}      
\end{figure}
A more detailed discussion that is summarized Figs. \ref{Figure1} and \ref{Figure2}. 
In Fig. \ref{Figure1} all the parameters are fixed except $n_{i}$ and $\alpha$.  
Along each of the curves $\log{r_{T}}$ is constant and the  labels refer to the value of the common logarithm (i.e. to base $10$) 
of $r_{T}$ computed from Eq. (\ref{rT1}). For illustration the arrows indicates the curve $\log{r_{T}} = -1$: the physical 
region, compatible with the current constraints, demands that $\log{r_{T}} < -1$. 

In Fig. \ref{Figure1} (plot on the left) the total number 
of efolds is $N_{t}=60$, while in the plot on the right $N_{t}=65$. As the number of efolds increases the value of $\alpha$ is pushed 
towards $0.1$.

The same trend is observed in Fig. \ref{Figure2} (plot on the left) where $r_{T}$ is illustrated in the $(\alpha, N_{t})$ plane.
As the total number of efolds increases beyond $65$, $\alpha$ is driven towards $0$. 

Whenever $n_{i}$ gets smaller than $1$ the parameter space in the $(\alpha,\, n_{1})$ plane gets larger, depending on the value of $n_{i}$. This region has been excluded since it would lead to a superluminal phase velocity which coincides, in this
case, with the group velocity. In specific situations where the group velocity does not coincide with the 
phase velocity, the regions $0<n_{i}<1$ might become phenomenologically viable. However in the present context we just 
want to focus on the most conservative situation.  
\begin{figure}[!ht]
\centering
\includegraphics[height=8cm]{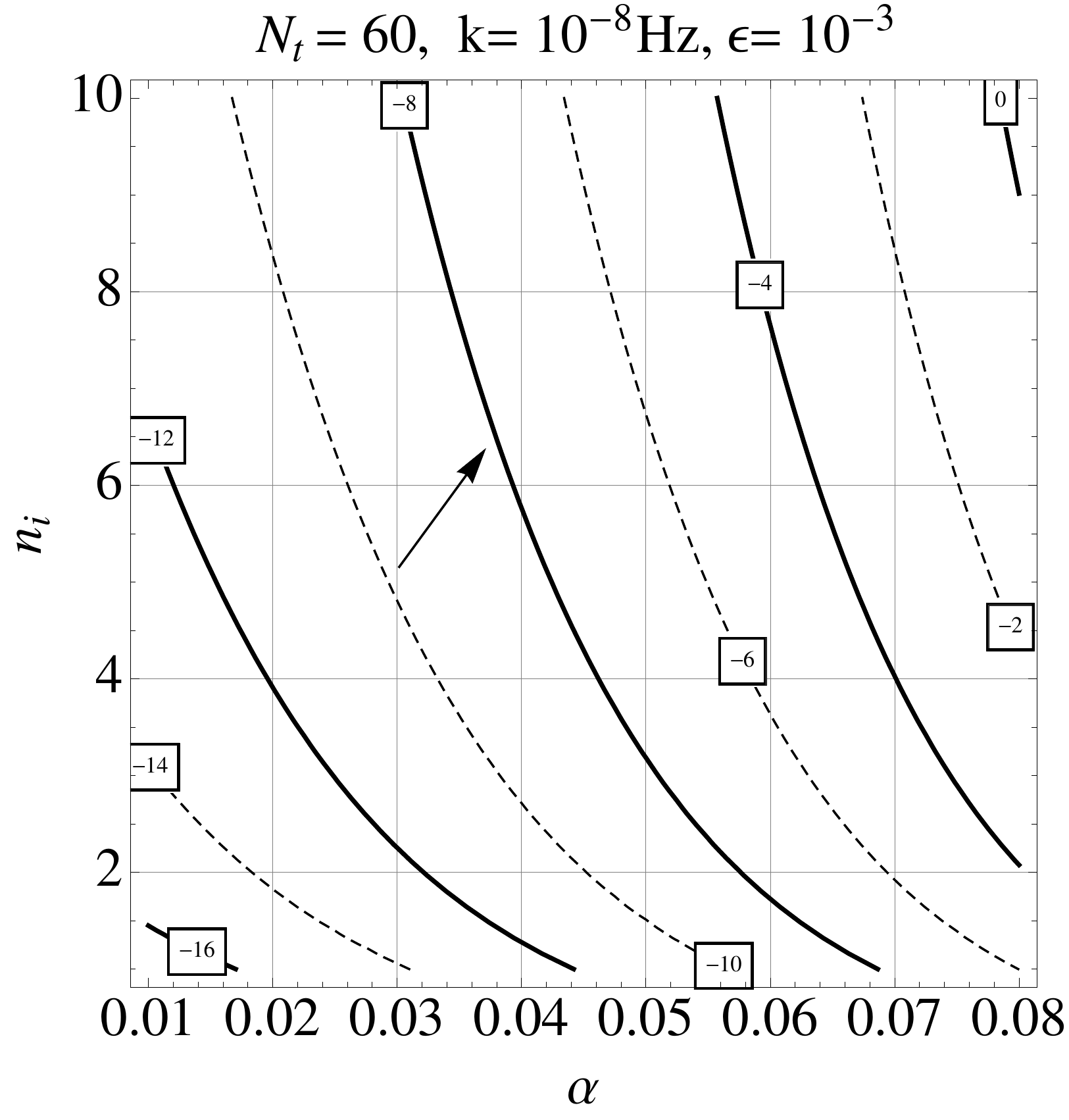}
\includegraphics[height=8cm]{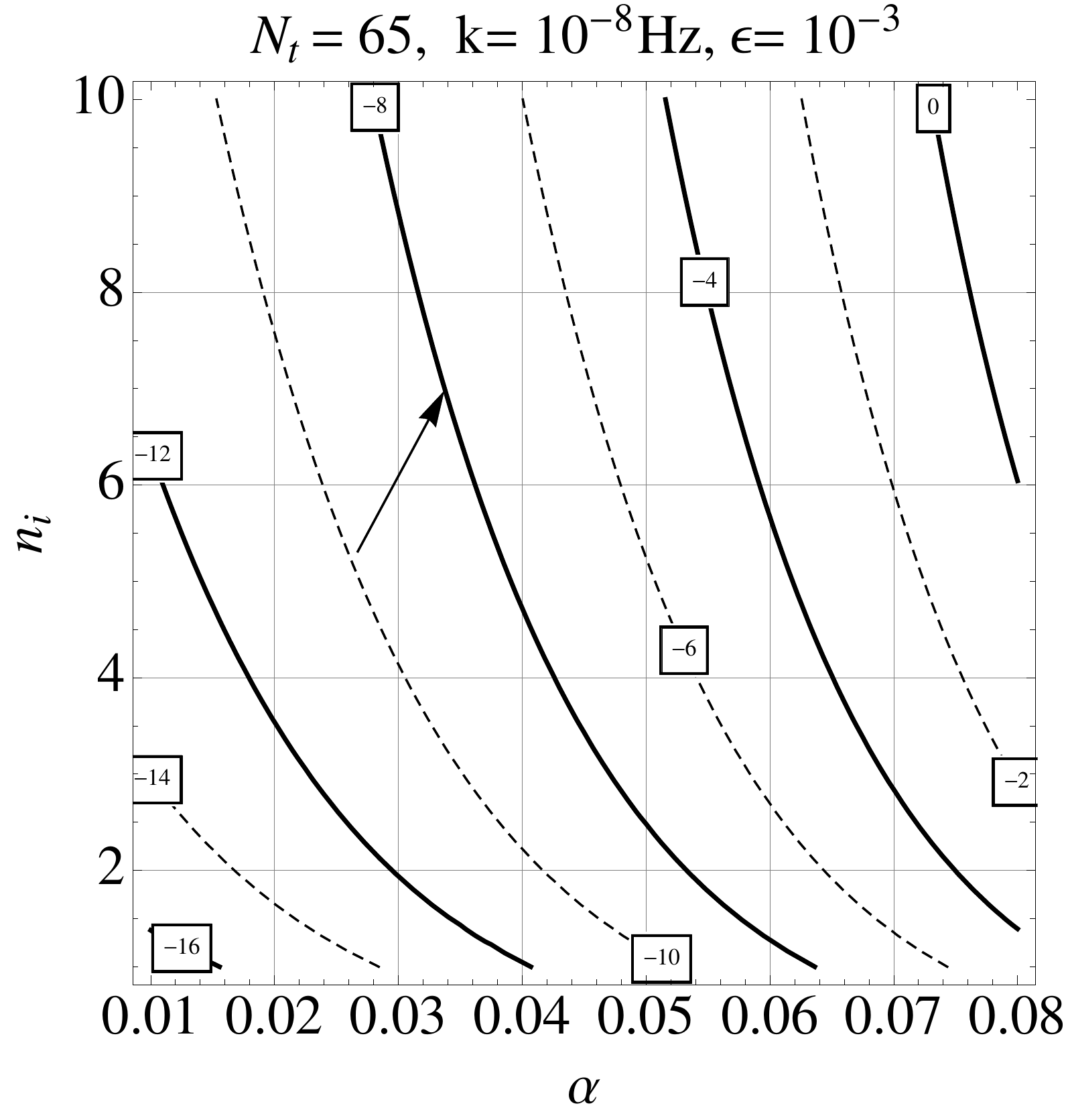}
\caption[a]{The value of the common logarithm $\Omega_{gw}$ computed from Eq. (\ref{OMFIN}) is illustrated in the $(\alpha, \, n_{i})$ plane and at the pulsar frequency.}
\label{Figure3}      
\end{figure}

\subsection{Limits from the pulsar timing bound}
\begin{figure}[!ht]
\centering
\includegraphics[height=8cm]{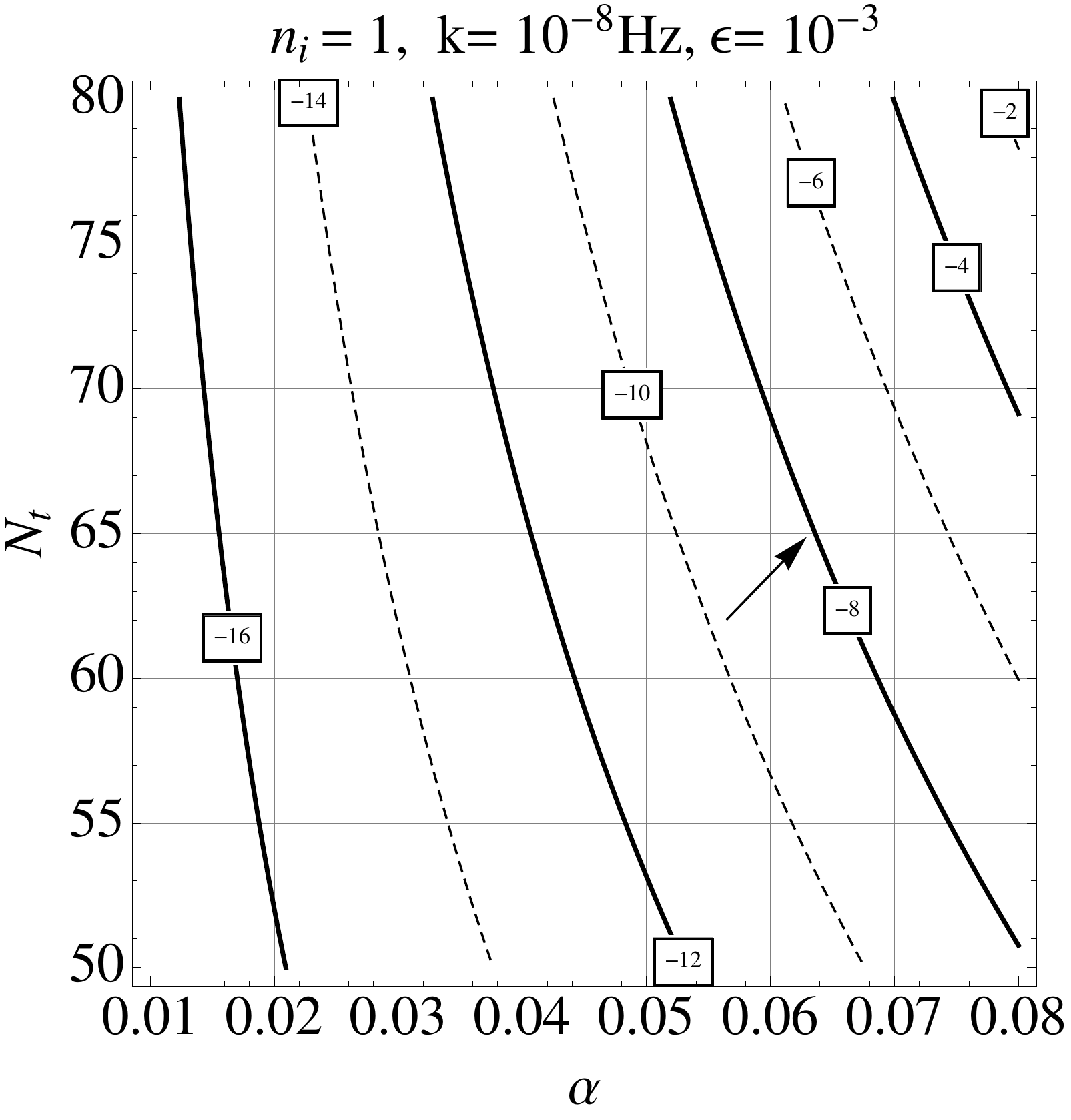}
\includegraphics[height=8cm]{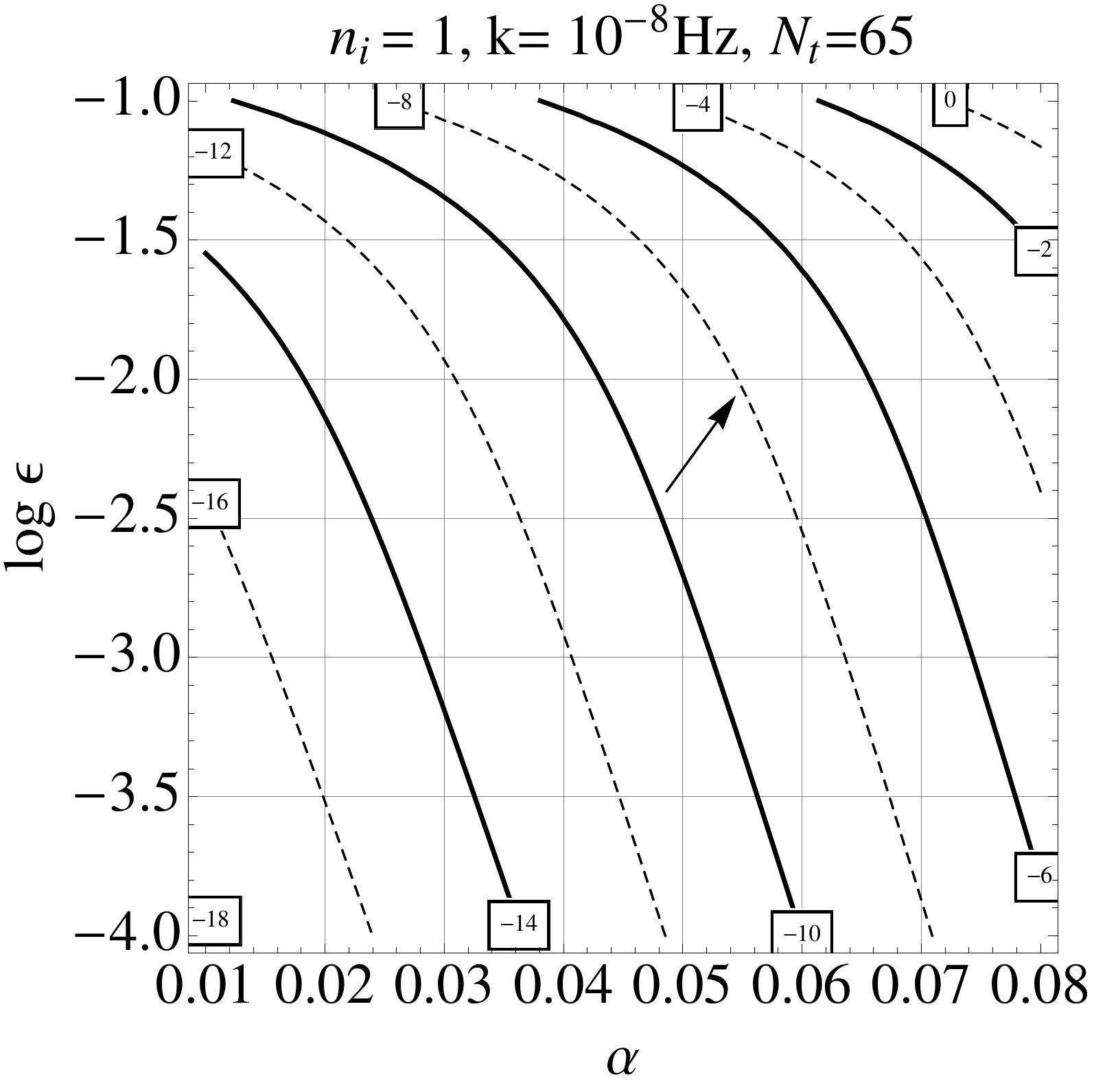}
\caption[a]{The common logarithm of $\Omega_{gw}$ at the pulsar frequency is illustrated in the $(\alpha, \, N_{t})$ plane and in the $(\alpha, \log{\epsilon})$ plane.}
\label{Figure4}      
\end{figure}

The bounds on $r_{T}$ are derived in the hypothesis that 
the consistency relations between the tensor amplitude and the tensor spectral index are verified. This is not necessarily true 
in the present context. Thus the bounds stemming from $r_{T}$ might be even less stringent than the ones we just analyzed. Ultimately this is not an important limitation since the most constraining bounds, in the present situation, come from higher wavenumbers (or higher  frequencies). Indeed, the pulsar timing constraint demands
\begin{equation}
\Omega(k_{pulsar},\tau_{0}) < 1.9\times10^{-8},\qquad 
k_{pulsar} \simeq \,10^{-8}\,\mathrm{Hz},
\label{PUL}
\end{equation}
where $k_{pulsar}$ roughly corresponds to the inverse 
of the observation time along which the pulsars timing has been monitored. The same strategy 
discussed in the previous subsection can now be applied to Eq. (\ref{PUL}). 
In Figs. \ref{Figure3} and \ref{Figure4} we used exactly the same range of parameters already 
employed in Figs. \ref{Figure1} and \ref{Figure2}.

In Figs. \ref{Figure3} and \ref{Figure4} we illustrate the common logarithm of $\Omega_{gw}$ computed from Eq. (\ref{OMFIN}). 
Comparing Figs. \ref{Figure3} with Fig. \ref{Figure1} the values of $\alpha$ allowed by the pulsar bound are much smaller than $0.1$ for the same range of variation of $n_{i}$. The same conclusion follows from the comparison of Figs. \ref{Figure4} and \ref{Figure2}. 
The pulsar bound is systematically more constraining because when the refractive index is dynamical the energy density of the relic gravitons is increasing (rather than decreasing) for $k \gg k_{eq}$. 
In Figs. \ref{Figure3} and \ref{Figure4} the arrow indicates, approximately, the curve where the pulsar bound is saturated. 

\subsection{Limits from the big-bang nucleosynthesis}
\begin{figure}[!ht]
\centering
\includegraphics[height=8cm]{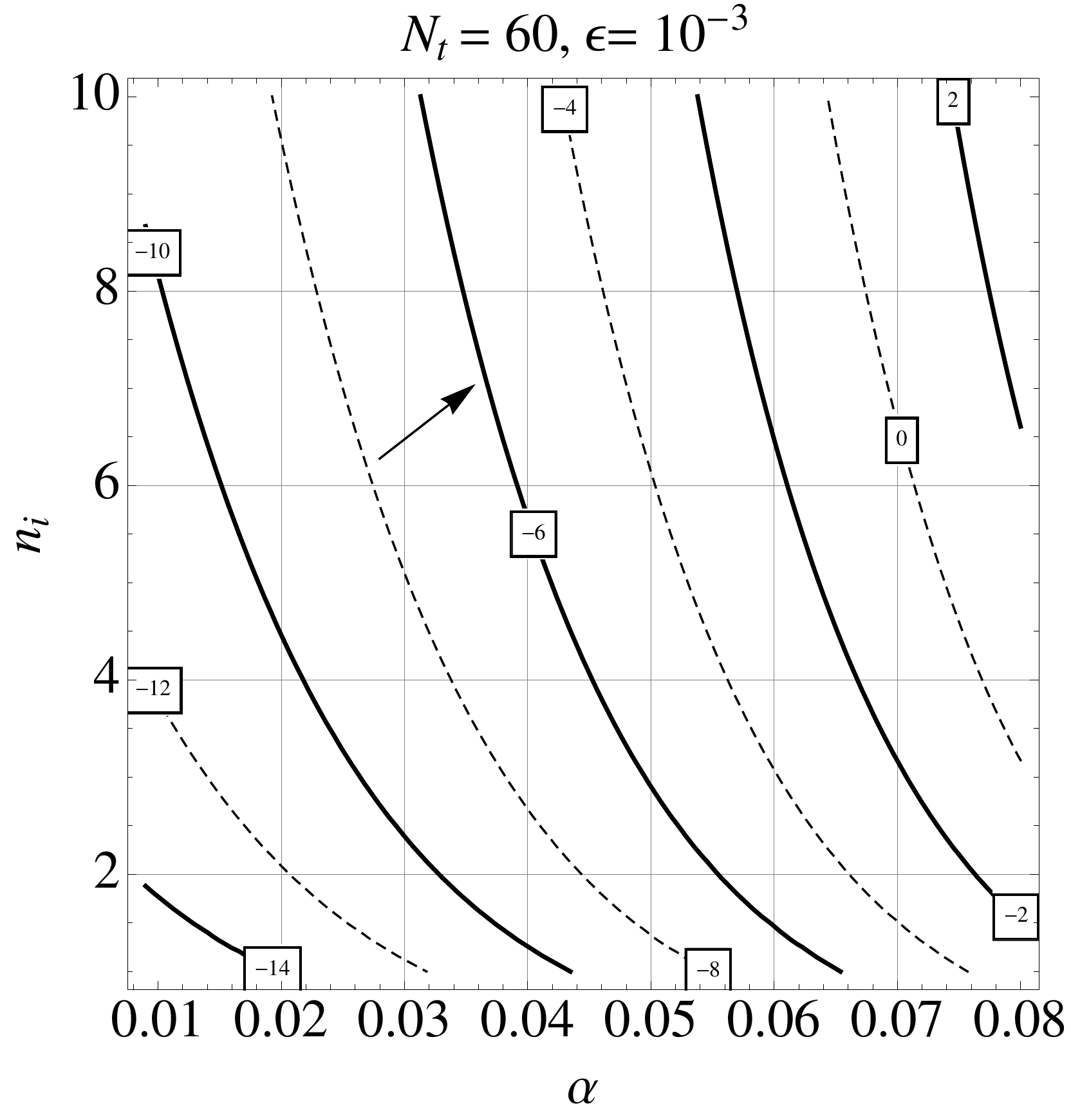}
\includegraphics[height=8cm]{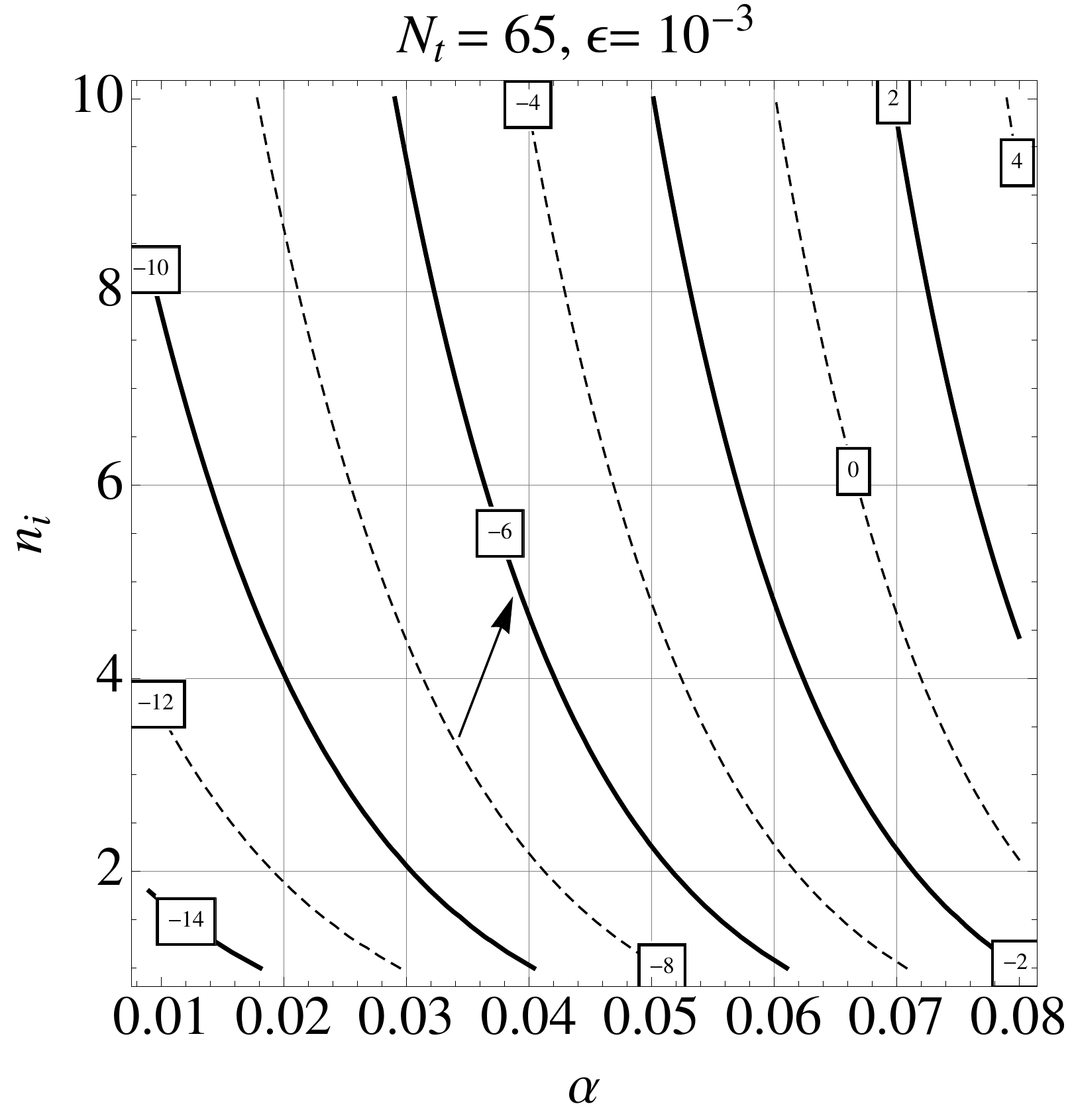}
\caption[a]{The common logarithm of the big-bang nucleosynthesis bound computed from Eq. (\ref{BBN1}) is illustrated in the $(\alpha, \, n_{i})$ plane.}
\label{Figure5}      
\end{figure}

\begin{figure}[!ht]
\centering
\includegraphics[height=8cm]{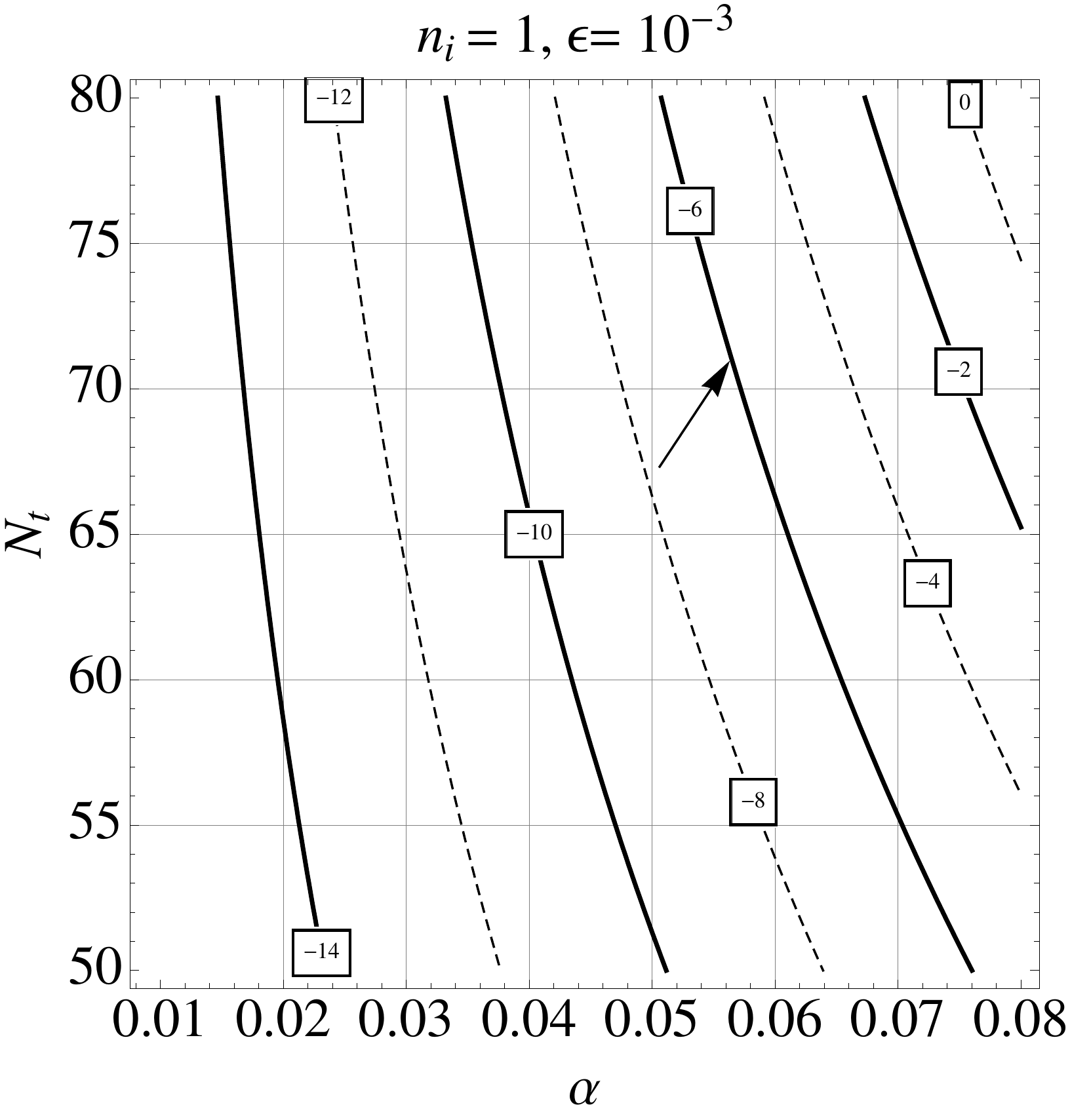}
\includegraphics[height=8cm]{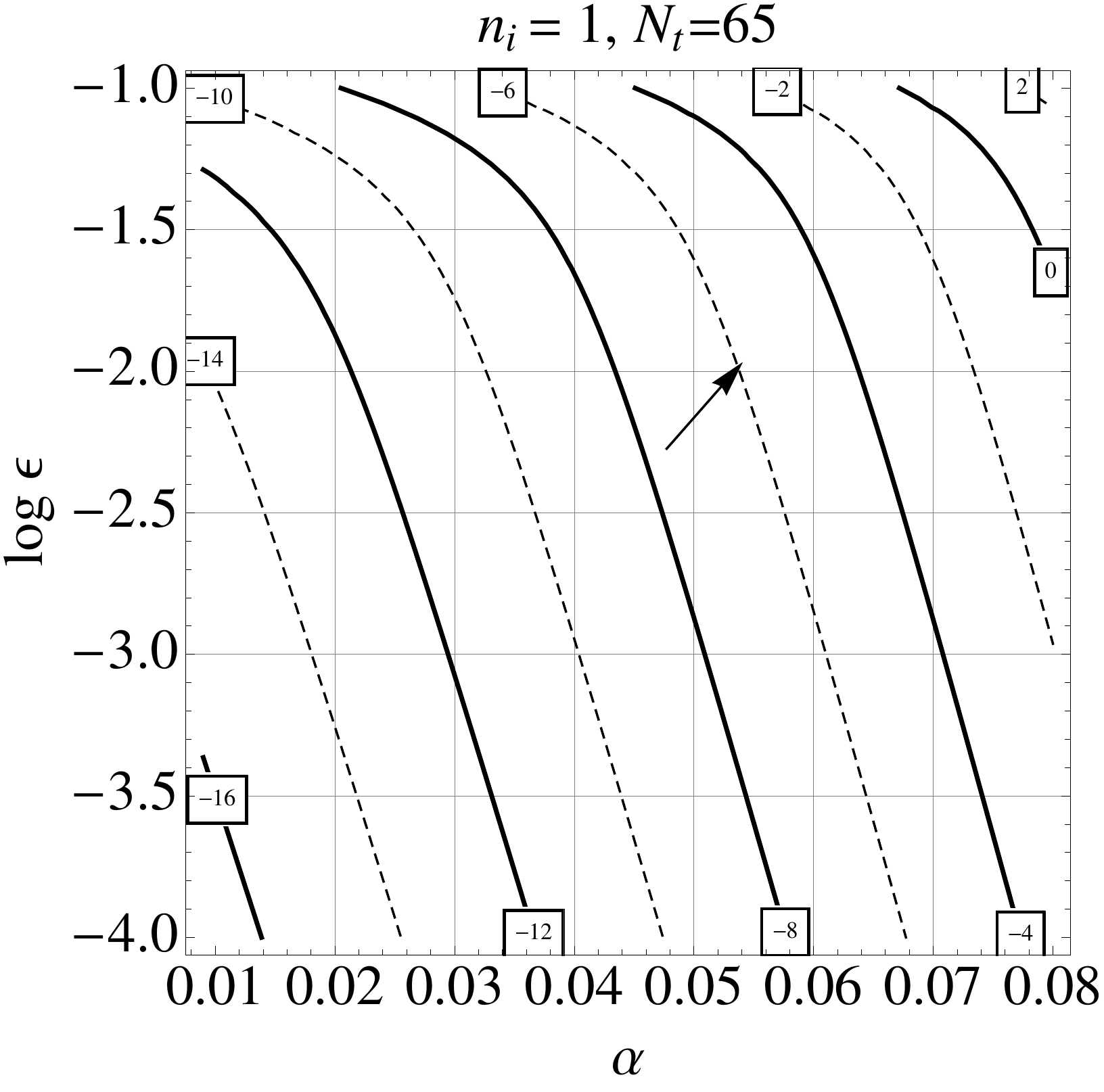}
\caption[a]{The common logarithm of the big-bang nucleosynthesis bound computed from Eq. (\ref{BBN1}) is illustrated in the  $(\alpha, \, N_{t})$ plane and in the $(\alpha, \log{\epsilon})$ plane.}
\label{Figure6}      
\end{figure}

A conclusion compatible with the pulsar bound can be drawn in the case of the big-bang nucleosynthesis constraint (BBN in what follows) stipulating that the bound on the extra-relativistic species at the time of big-bang nucleosynthesis can be 
translated into a bound on the cosmic graviton backgrounds. 
This constraint is customarily expressed in terms of $\Delta N_{\nu}$ representing the contribution of supplementary 
neutrino species but the extra-relativistic species do not need to be fermionic. If the additional species are 
relic gravitons we have: 
\begin{equation}
h_{0}^2  \int_{k_{bbn}}^{k_{max}}
  \Omega_{{\rm GW}}(k,\tau_{0}) d\ln{k} = 5.61 \times 10^{-6} \Delta N_{\nu} 
  \biggl(\frac{h_{0}^2 \Omega_{\gamma0}}{2.47 \times 10^{-5}}\biggr),
\label{BBN1}
\end{equation}
where $k_{bbn}$ and $k_{max}$ have been computed, respectively, in Eqs. (\ref{kbbn}) and (\ref{kmax}); note that in Eq. (\ref{BBN1}) 
$\Omega_{\gamma 0}$ denotes the present critical fraction of energy density coming just from photons. 
The bounds on $\Delta N_{\nu}$ range from $\Delta N_{\nu} \leq 0.2$ 
to $\Delta N_{\nu} \leq 1$.
The bounds stemming from Eq. (\ref{BBN1}) can be easily inferred from Figs. \ref{Figure5} and \ref{Figure6}. In both 
cases we illustrate the common logarithm of the left hand side of Eq. (\ref{BBN1}) computed in the case of a dynamical refractive 
index from Eq. (\ref{OMFIN}).
By comparing Figs. \ref{Figure5} and \ref{Figure3} we see that the pulsar and the BBN bound are largely compatible. Conversely 
by comparing Figs. \ref{Figure5} and \ref{Figure1} the big-bang nucleosynthesis bound is 
always more constraining than the bounds stemming from $r_{T}(k_{p})$. 
The same conclusion is reached if we compare the variation of a different set of parameters, like 
 in Figs. \ref{Figure2}, \ref{Figure4} and \ref{Figure6}. Note finally that 
in Figs. \ref{Figure5} and \ref{Figure6} the arrow has been used to guide the eye towards the curve 
where the big-bang nucleosynthesis bound is approximately saturated. 

\subsection{Prospects for wide-band detectors}
The bounds examined in the previous subsections suggest that for 
the standard fiducial values of $(N_{t},\, \epsilon)$ the values of $n_{i}$ and $\alpha$ are constrained to be within the following window:
\begin{equation}
1< n_{1} < 10, \qquad 0 <\alpha < 0.07.
\label{lim1}
\end{equation}
The sensitivity of a given pair of wide-band detectors to a stochastic background of relic gravitons depends upon 
the relative orientation of the instruments (see e.g. \cite{corr,corr2}) and a specific analysis of the signal to noise 
ratio is beyond the scopes of this paper. While the advanced version of the wide-band interferometers is still matter 
of debate (and the published results do not seem conclusive at the moment) the frequency window of the detectors will always be between few Hz (where the seismic noise 
dominates) and $10$ kHz (where the shot noise eventually dominates). The wideness of the band 
(important for the correlation among different instruments)
is not as large as $10$ kHz  but much narrower.  There are projects of wide-band detectors in space like the Lisa, the Bbo/Decigo. The common feature of these three projects is that they are all space-borne missions; the Lisa interferometer should operate between $10^{-4}$ and $0.1$ Hz. Nominally the Decigo project will be instead sensitive to frequencies between $0.1$ and $10$ Hz.

Recalling the results obtained so far we can safely say that growing spectra can arise in the following range:
\begin{equation}
\frac{2\epsilon}{( 3 + 5 \epsilon)} < \alpha <  - \frac{\ln{(\pi \, \epsilon \, {\mathcal A}_{{\mathcal R}})}}{ 3\, N_{t}}, 
\label{gr}
\end{equation}
where the lower bound of Eq. (\ref{gr}) has been derived after Eq. (\ref{inf8}) while the upper bound follows from Eq. (\ref{endens3}). Since the the upper limit of Eq. (\ref{gr}) is larger than the one of Eq. (\ref{lim1}) we can conclude 
that growing spectra arise in practice in the whole range of variation of $\alpha$. 
Using the illustrative example discussed before we have that for $k_{Lisa}={\mathcal O}(10^{-3})$ Hz (i.e. compatible 
with the Lisa window) we would have $\Omega_{\mathrm{gw}}(\mathrm{mHz}) \simeq 10^{-8.21}$
for $\alpha =0.06$, $N_{t} =65$, $\epsilon =0.001$ and $n_{i} =1$. Larger values of $n_{i}$ and $N_{t}$ make the signal smaller. 
For the Ligo/Virgo frequencies and for the same parameters chosen in the Lisa case  we would have instead $\Omega_{\mathrm{gw}}(0.1\, \mathrm{kHz}) \simeq 10^{-7.69}$ where $k_{Ligo}=0.1$ kHz. 

\begin{figure}[!ht]
\centering
\includegraphics[height=8cm]{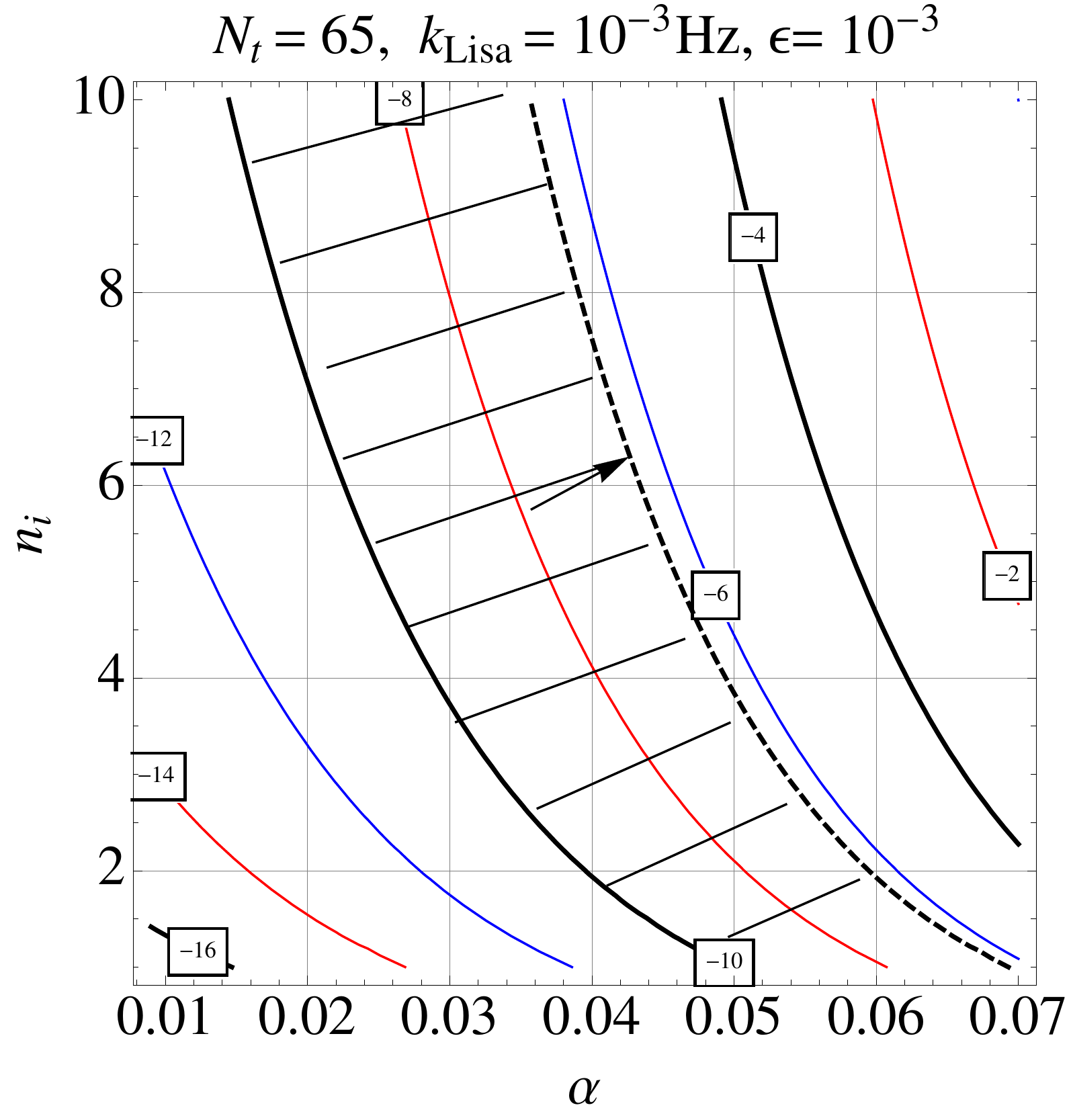}
\includegraphics[height=8cm]{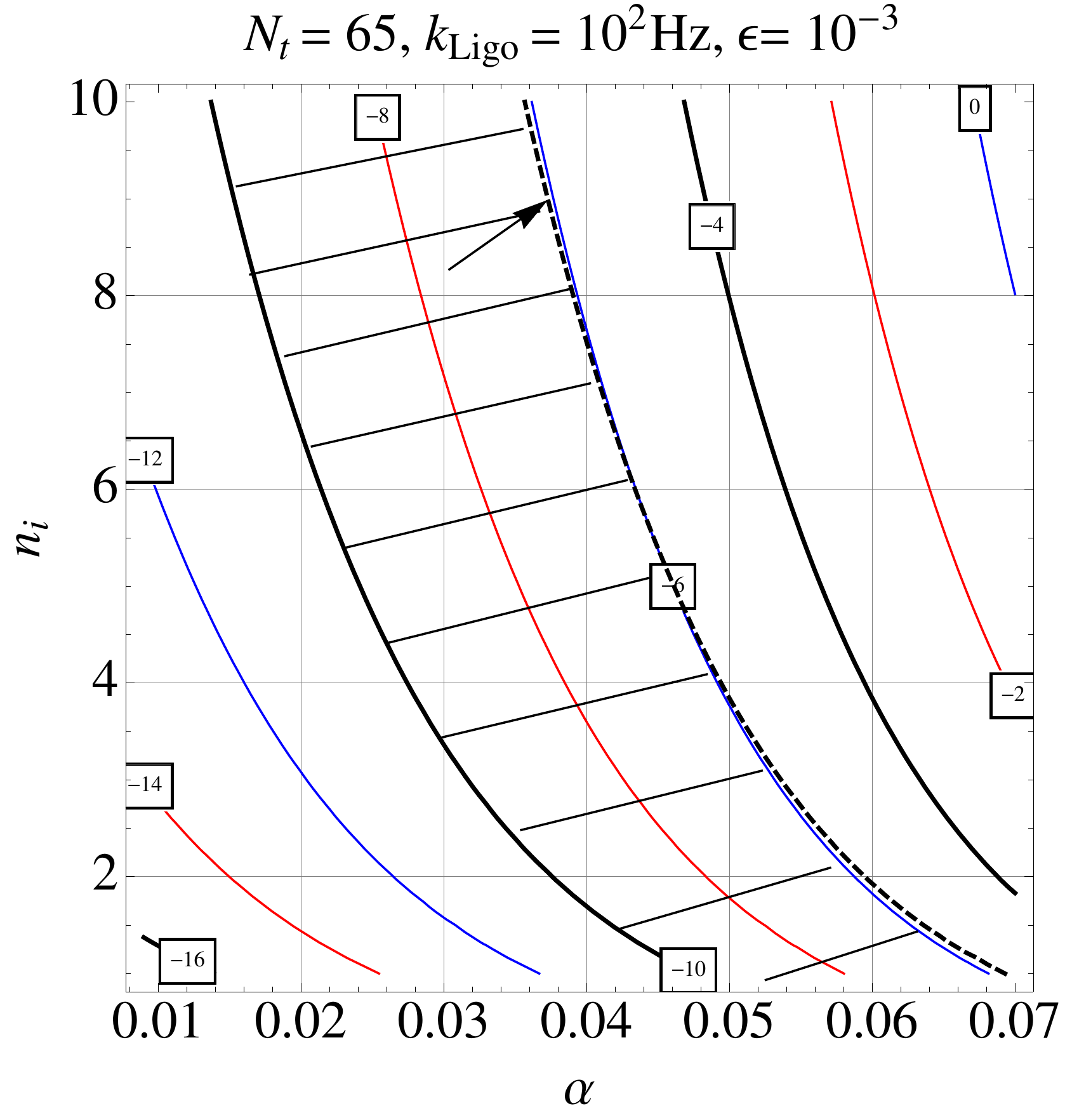}
\caption[a]{The energy density of the relic gravitons is illustrated in the  $(\alpha, \, n_{i})$ plane in the Lisa window (plot on the left) and in the Ligo/Virgo window (plot on the right).}
\label{Figure7}      
\end{figure}

This trend is confirmed by the results illustrated in Fig. \ref{Figure7}
where we report the common logarithm of $\Omega_{gw}$ in the Lisa window (plot on the left) and in the 
Ligo/Virgo window (plot on the right). The dashed lines in both plots corresponds to the common logarithm of the big-bang nucleosynthesis 
bound in the $\Delta N_{\nu} =1$ case (i.e. $\log{(5.61\times 10^{-6})} = - 5.25$). 
The allowed region of the parameter space must be, in both plots of Fig. \ref{Figure7}, 
below the dashed lines\footnote{By comparing the two plots of Fig. \ref{Figure7} the dashed line (representing the big-bang 
nucleosynthesis bound) looks closer to the actual value of the $\log{\Omega_{gw}}$ in the Ligo/Virgo case than in the Lisa case. Indeed, if the integrand increases,  the integral appearing in Eq. (\ref{BBN1}) can be approximated by the value of $\Omega_{gw}$ at $k_{max}$.}. While the noise 
power spectra of Lisa are still rather hypothetical, the advanced version of terrestrial interferometers 
might get down to $10^{-10}$ in $\Omega_{gw}$ after an appropriate integration time. The shaded area in Fig. \ref{Figure7} 
illustrates  the allowed region in the parameter space where the relic gravitons are potentially detectable. 
As already mentioned we shall not dwell here on the detectability prospects in this paper. The value $\Omega_{gw} = {\mathcal O}(10^{-10})$
has been mainly quoted to guide the eye and whether this sensitivity will be in fact reached by terrestrial of space-borne
detectors is an entirely different issue. If this value will be indeed reached by cross-correlation of two instruments more detailed 
discussions could be necessary (see e.g. \cite{corr,corr2} and discussions therein). 

It is finally appropriate to mention that 
the maximal signal due to the variation of the refractive index 
occurs in a frequency region between the MHz and the GHz. 
In this range of frequencies microwave cavities or even wave guides 
\cite{HF1,HF2,HF3} can be used as detectors of gravitational waves.

\renewcommand{\theequation}{5.\arabic{equation}}
\setcounter{equation}{0}
\section{Concluding remarks}
\label{sec5}
The continuous and differentiable evolution of the refractive index leads 
to power spectra and spectral energy densities of the relic gravitons that are slightly increasing as a function of the comoving wavenumber 
(or of the comoving frequency). While the rate of variation of the refractive index 
can be stringently bounded, the derived limits do not exclude  
the potential relevance of the cosmic graviton background for either ground based or space-borne interferometers
aimed at a direct detection of gravitational waves.
The spectral slopes are determined by the competition of the slow-roll parameter against $\alpha$ which measures the rate of variation 
of the refractive index in units of the Hubble rate. The phenomenologically 
allowed region implies that $1\leq n_{i} <10$ and $0<\alpha < 0.07$, where $n_{i}$ denotes the value 
of the refractive index at the onset of the inflationary expansion, 

The sensitivity of wide-band detectors\footnote{We recall, for the sake of precision, that the expression of the signal-to-noise ratio 
 in the context of optimal processing  required for the detection of stochastic backgrounds \cite{corr,corr2} depends on an integral over the frequency band (between few Hz and 10 kHz in the case of the ground-based interferometers of Ligo-type). The numerator of the integrand 
 contains $\Omega_{gw}^2$ while the denominator we have the sixth power of the frequency multiplied the noise power spectra of each of the 
 two correlated detectors. The signal-to-noise ratio depends also on the relative orientation of the interferometers and on the total 
 observation time which is crucial to increase the sensitivity. Naively, if the minimal detectable signal (by one detector) is 
 $h_{0}^2\Omega_{gw}$, then the cross-correlation of two 
 identical instruments might increase the sensitivity by a factor $1/\sqrt{\Delta \nu T}$ where 
 $\Delta \nu$ is the bandwidth and $T$, as already mentioned, is the observation time.
Therefore if a single instrument detects  $h_{0}^2\Omega_{gw} \simeq 10^{-5}$ the correlation may detect
 $h_{0}^2\Omega_{\mathrm{GW}} \simeq 10^{-10}$ provided $\Delta \nu \simeq 100$ Hz and 
 $T\simeq {\mathcal O}(1\mathrm{yr})$ \cite{max12}.}  to the relic graviton backgrounds is customarily 
expressed in terms of the minimal detectable spectral energy density of the relic gravitons in critical units.
The operating windows of the ground based 
and space-borne interferometers are complementary: the typical frequency range of space-borne interferometers 
extends between $0.1 \,\,\mathrm{mHz} = 10^{-4}\, \mathrm{Hz}$ and few Hz. Conversely the window of ground based 
detectors extends between few Hz (where seismic noise dominates) and $10\, \mathrm{kHz} = 10^{4} \, \mathrm{Hz}$ (where shot noise 
dominates). While the time scales for the realization of space-borne interferometers are still vague
 it is useful to illustrate our findings by keeping ground based detectors and space-borne interferometers on equal footing. 

In Tab. \ref{Table1} the frequencies encompass the operating ranges of space-borne and ground-based 
detectors. In the first column we illustrate the common logarithm of $\Omega_{gw}$. In the two remaining columns we illustrate 
the common logarithms of the strain amplitude ${\mathcal S}_{h}$ and of its square root\footnote{The sensitivity is often expressed by means of  ${\mathcal S}_{h}$ or in terms of its square root\cite{max12}. The precise relation between ${\mathcal S}_{h}$ and $\Omega_{gw}$ is given by  ${\mathcal S}_{h}(\nu,\tau_{0}) = 7.981\times 10^{-43} \,\,(100\,\mathrm{Hz}/\nu)^3 \,\, h_{0}^2 \Omega_{gw}(\nu,\tau_{0})\,\, \mathrm{Hz}^{-1}$. Note that ${\mathcal S}_{h}$ is measured in $1/\mathrm{Hz}= \mathrm{sec}$.}.
\begin{table}
\begin{center}
\vskip 0.5truecm
\begin{tabular}{| c | | | c | l | c | | | c | | |  c | }
\hline
 $\nu$ \quad & $\log{[\Omega_{gw}]}$ & $\log{[{\mathcal S}_{h} \mathrm{Hz}]} $ & $\log{[ \sqrt{{\mathcal S}_{h} \,\mathrm{Hz}}]}$   \\ \hline
$10^{-4} \,\,\mathrm{Hz}$ &  $-8.15$ &       $-32.56$                 &  $-16.28$ \\ \hline
$10^{-2}\,\,\mathrm{Hz}$ &  $-7.94$ &       $-38.35$                  &  $-19.17$\\ \hline
$1\,\, \,\,\mathrm{Hz}$ &       $-7.74$ &       $-44.14$                 &  $-22.07$\\ \hline
$10^{2}\,\, \mathrm{Hz}$ &  $-7.53$ &       $-49.93$                  &  $-24.96$\\ \hline
$10^{4}\,\, \mathrm{Hz}$ &  $-7.32$ &       $-55.73$                  &  $-27.86$\\ \hline
$10^{6}\,\, \mathrm{Hz}$ &  $-7.11$ &       $-61.52$                  &  $-30.76$\\ \hline
$10^{8}\,\, \mathrm{Hz}$ &  $-6.90$ &       $-67.31$                  &  $-33.65$\\ \hline
$10^{10}\,\, \mathrm{Hz}$ &  $-6.69$ &       $-73.10$                &  $-22.07$\\ \hline
\hline
\end{tabular}
\caption{Typical values of the spectral energy density and of the strain amplitude in the operating windows of ground-based and space-borne interferometers. Consistently with the discussion of section \ref{sec4} the parameters  $(n_{i}, \alpha, \epsilon, N_{t})$ have been chosen to be $(1,\,0.06,\, 0.001,\,65)$.}
\label{Table1}
\end{center}
\end{table}
In Tab. \ref{Table1} the range of frequencies has been extended well beyond the window of ground-based interferometers. The last 
frequency, i.e. $10\, \mathrm{GHz}$ is even larger than $\nu_{max} = k_{max}/(2\pi) = {\mathcal O}(\mathrm{GHz})$ and we just included it to show that the spectral energy density is always well below the constraints previously discussed. 

If the correlation of (advanced) wide-band interferometers 
will eventually reach sensitivities ${\mathcal O}(10^{-10})$ in $\Omega_{gw}$, the present considerations might become 
relevant since the signal due to a dynamical refractive index  with $\alpha = {\mathcal O}(0.06)$ and $n_{i} = {\mathcal O}(1)$ 
can even be ${\mathcal O}(10^{-7.53})$ for the phenomenologically 
allowed region of the parameter space and for a typical frequency ${\mathcal O}(0.1 \, \mathrm{kHz})$. 
Even if the noise power spectra and the specific features of the
hypothetical space-borne interferometers  are still unclear, a potential sensitivity ${\mathcal O}(10^{-8.05})$ in $\Omega_{gw}$ cannot be excluded for a typical frequency ${\mathcal O}(\mathrm{mHz})$. 
The lack of detection of a stochastic background of relic gravitons either by ground based detectors or by space-borne interferometers 
may therefore provide a further (and potentially much more stringent) bound on the rate 
of variation of the refractive index. 

\newpage

\begin{appendix}
\renewcommand{\theequation}{A.\arabic{equation}}
\setcounter{equation}{0}
\section{Mixing coefficients during the radiation phase}
\label{APPA}
The explicit expression of the mixing coefficients appearing in Eq. (\ref{rad1}) can be written as
\begin{eqnarray}
c_{+}(k,\tau_{1}) &=& - \frac{\pi \, i}{4 ( 1 - \alpha)} \biggl[ \sqrt{\beta} \,q \, H_{\mu}^{(1)}(\overline{g}_{i}) \,{\mathcal A}^{(1)}_{\rho}(\overline{g}_{r}) 
- \frac{q}{\sqrt{\beta}}  H_{\rho}^{(1)}(\overline{g}_{r})\, {\mathcal B}^{(1)}_{\mu}(\overline{g}_{i}) \biggr],
\label{cp1}\\
c_{-}(k,\tau_{1}) &=& \frac{\pi \, i}{4 ( 1 - \alpha)} \biggl[ \sqrt{\beta} \,q \,H_{\mu}^{(1)}(\overline{g}_{i}) \,{\mathcal A}^{(2)}_{\rho}(\overline{g}_{r}) 
+ \frac{q}{\sqrt{\beta}}  H_{\rho}^{(2)}(\overline{g}_{r}) \, {\mathcal B}^{(1)}_{\mu}(\overline{g}_{i}) \biggr],
\label{cm1}
\end{eqnarray}
where $q = \sqrt{|1-\alpha|/|1 + \alpha\beta|}$. In Eqs. (\ref{cp1}) and (\ref{cm1}) the following notation have been employed: 
\begin{equation}
g_{i}(-\tau_{1})=\overline{g}_{i}= \frac{k \tau_{1}}{n_{1} | 1 + \alpha \beta|}, \qquad  g_{r}(-\tau_{1})=\overline{g}_{r}= \frac{k \tau_{1}}{n_{1} | 1 - \alpha| \beta}.
\label{gg1}
\end{equation}
Furthermore, in Eqs. (\ref{cp1}) and (\ref{cm1})  ${\mathcal A}_{\rho}^{(1)}(\overline{g}_{r})$ and  ${\mathcal B}_{\mu}^{(1)}(\overline{g}_{i}) $ 
are two auxiliary expressions defined as:
\begin{eqnarray}
{\mathcal A}_{\rho}^{(1)}(\overline{g}_{r}) &=& H_{\rho}^{(1)}(\overline{g}_{r}) \biggl[ (1-\alpha)\rho + \frac{1}{2}\biggr] - \overline{g}_{r} (1 - \alpha) H_{\rho+1}^{(1)}(\overline{g}_{r}),
\label{AA1}\\
{\mathcal B}_{\mu}^{(1)}(\overline{g}_{i}) &=& H_{\mu}^{(1)}(\overline{g}_{i}) \biggl[ (1+\alpha\beta)\mu + \frac{1}{2}\biggr] - \overline{g}_{i} (1 + \alpha\beta) H_{\mu+1}^{(1)}(\overline{g}_{i}),
\label{AA2}
\end{eqnarray}
where, following the properties of the Hankel functions under complex conjugation, we shall have 
${\mathcal A}_{\rho}^{(2)}(\overline{g}_{r}) = {\mathcal A}_{\rho}^{(1)\,\ast}(\overline{g}_{r})$ and 
${\mathcal A}_{\mu}^{(2)}(\overline{g}_{i}) = {\mathcal A}_{\mu}^{(1)\,\ast}(\overline{g}_{i})$.
In more explicit terms Eqs. (\ref{cp1}) and (\ref{cm1}) can be written as:
\begin{eqnarray}
c_{+}(k,\tau_{1}) &=& - \frac{i \pi \sqrt{\beta}}{4 ( 1 - \alpha)} \sqrt{\frac{|1-\alpha|}{|1 + \alpha \beta|} }\biggl\{H_{\mu}^{(1)}(\overline{g}_{i}) \, H_{\rho}^{(1)}(\overline{g}_{r}) \biggl[ \frac{\beta + 1 }{ 2 \beta} + (1 - \alpha) \rho + \frac{\mu(1 + \alpha \beta)}{\beta}\biggr]
\nonumber\\
&-& (1 - \alpha) \overline{g}_{r} \, H_{\mu}^{(1)}(\overline{g}_{i}) H_{\rho+1}^{(1)}(\overline{g}_{r}) - \frac{(1 + \alpha \beta)}{\beta} \,\overline{g}_{i} \, H_{\rho}^{(1)}(\overline{g}_{r}) \,H_{\mu + 1}^{(1)}(\overline{g}_{i})\biggr\},
\label{cp2}\\
c_{-}(k,\tau_{1}) &=& \frac{i \pi \sqrt{\beta}}{4 ( 1 - \alpha)} \sqrt{\frac{|1-\alpha|}{|1 + \alpha \beta|} }\biggl\{H_{\mu}^{(1)}(\overline{g}_{i}) \, H_{\rho}^{(2)}(\overline{g}_{r}) \biggl[ \frac{\beta + 1 }{ 2 \beta} + (1 - \alpha) \rho + \frac{\mu(1 + \alpha \beta)}{\beta}\biggr]
\nonumber\\
&-& (1 - \alpha) \overline{g}_{r} \, H_{\mu}^{(1)}(\overline{g}_{i}) H_{\rho+1}^{(2)}(\overline{g}_{r}) - \frac{(1 + \alpha \beta)}{\beta} \, \overline{g}_{i} \,H_{\rho}^{(2)}(\overline{g}_{r}) \,H_{\mu + 1}^{(1)}(\overline{g}_{i})\biggr\}.
\label{cm2}
\end{eqnarray}
As it can be explicitly verified from Eqs. (\ref{cp2}) and (\ref{cm2}), $|c_{+}(k,\tau_{1})|^2 - |c_{-}(k,\tau_{1})|^2 =1$. 
It is useful to mention that Eqs. (\ref{cp2}) and (\ref{cm2}) reproduce exactly the standard results in the limit $\alpha\to 0$, $n_{1}\to 1$  
and $\beta\to 1$. More specifically this limit refers to the situation where there is a transition from an exact de Sitter phase (i.e. $\epsilon \to 0$ and $\mu \to 3/2$) to 
a conventional radiation-dominated phase where $\rho \to 1/2$. In the standard limit we also have that 
$\overline{g}_{i} \to k \tau_{1}$ and $ \overline{g}_{r} \to k \tau_{1}$ and, from the explicit form of Eqs. (\ref{cp2}) and (\ref{cm2}),
\begin{equation}
c_{+}( x_{1}) = \frac{i}{2\,x_{1}^2} [ 2 x_{1} ( x_{1} + i) -1] e^{ 2 i x_{1}}, \qquad c_{-}( x_{1}) = - \frac{i}{2\,x_{1}^2};
\label{cpm}
\end{equation}
in this case the mixing coefficients depend on the single argument $x_{1}= k\tau_{1}$.

\renewcommand{\theequation}{B.\arabic{equation}}
\setcounter{equation}{0}
\section{Mixing coefficients during the matter phase}
\label{APPB}
The explicit expression of the mixing coefficients appearing in Eq. (\ref{mat1}) can be written as
\begin{eqnarray}
d_{+}(k,\tau_{1},\tau_{2}) &=& - \frac{i \pi}{4 \sqrt{2} ( 1 - 2 \alpha)}\sqrt{ \frac{|1-2\alpha|}{|1-\alpha|}}\biggl\{  c_{+}(k,\tau_{1})  \biggl[ H_{\rho}^{(2)}(\widetilde{g}_{r}) {\mathcal F}_{\sigma}^{(1)}(\widetilde{g}_{m}) - 2 H_{\sigma}^{(1)}(\widetilde{g}_{m}) {\mathcal G}_{\rho}^{(2)}(\widetilde{g}_{r})\biggr]
\nonumber\\
&+& 
c_{-}(k,\tau_{1}) \biggl[ H_{\rho}^{(1)}(\widetilde{g}_{r}) {\mathcal F}_{\sigma}^{(1)}(\widetilde{g}_{m}) - 2 H_{\sigma}^{(1)}(\widetilde{g}_{m}) {\mathcal G}_{\rho}^{(1)}(\widetilde{g}_{r})\biggr]\biggr\},
\label{dp1}\\
d_{-}(k,\tau_{1},\tau_{2}) &=&  \frac{i \pi}{4 \sqrt{2} ( 1 - 2 \alpha)}\sqrt{ \frac{|1-2\alpha|}{|1-\alpha|}}\biggl\{  c_{+}(k,\tau_{1})  \biggl[ H_{\rho}^{(2)}(\widetilde{g}_{r}) {\mathcal F}_{\sigma}^{(2)}(\widetilde{g}_{m}) - 2 H_{\sigma}^{(2)}(\widetilde{g}_{m}) {\mathcal G}_{\rho}^{(2)}(\widetilde{g}_{r})\biggr]
\nonumber\\
&+& c_{-}(k,\tau_{1}) \biggl[ H_{\rho}^{(1)}(\widetilde{g}_{r}) {\mathcal F}_{\sigma}^{(2)}(\widetilde{g}_{m}) - 2 H_{\sigma}^{(2)}(\widetilde{g}_{m}) {\mathcal G}_{\rho}^{(1)}(\widetilde{g}_{r})\biggr]\biggr\},
\label{dm1}
\end{eqnarray}
where $\widetilde{g}_{m}$ and $\widetilde{g}_{r}$ are defined as:
\begin{equation}
g_{m}(\tau_{2})=\widetilde{g}_{m} = \frac{2 k \tau_{2}}{n_{1}|1-2\alpha|} \biggl(\frac{\tau_{1}}{\beta \tau_{2}}\biggr)^{\alpha} \biggl[1 + \frac{\beta + 1}{\beta} \biggl(\frac{\tau_{1}}{\tau_{2}}\biggr) \biggr]^{ 1 - \alpha}, \qquad g_{r}(\tau_{2})= \widetilde{g}_{r} =  \frac{|1 - 2\alpha|}{2 | 1 - \alpha|} \widetilde{g}_{m}.
\label{dpm2}
\end{equation}
If Eqs. (\ref{dp1}) and (\ref{dm1}) the following auxiliary functions have been introduced:
\begin{eqnarray}
{\mathcal F}_{\sigma}^{(1)}(\widetilde{g}_{m})&=& \biggl[ \frac{1}{2} + \sigma ( 1 - 2 \alpha) \biggr] H_{\sigma}^{(1)}(\widetilde{g}_{m}) - ( 1 - 2 \alpha) \widetilde{g}_{m} \,H_{\sigma+1}^{(1)}(\widetilde{g}_{m}),
\label{F1}\\
 {\mathcal G}_{\rho}^{(2)}(\widetilde{g}_{r}) &=& \biggl[ \frac{1}{2} + \rho ( 1 -  \alpha) \biggr] H_{\rho}^{(1)}(\widetilde{g}_{r}) 
 - ( 1 -  \alpha) \widetilde{g}_{r}\, H_{\rho +1 }^{(1)}(\widetilde{g}_{r}),
\label{G1}
\end{eqnarray}
furthermore, following the standard notations, 
we have that ${\mathcal F}_{\sigma}^{(2)}(\widetilde{g}_{m})= {\mathcal F}_{\sigma}^{(1)\,\ast}(\widetilde{g}_{m})$ and 
${\mathcal G}_{\rho}^{(2)}(\widetilde{g}_{r})= {\mathcal G}_{\rho}^{(1)\,\ast}(\widetilde{g}_{r})$. Using Eqs. (\ref{F1}) and (\ref{G1}) 
the relation $|d_{+}|^2  - |d_{-}|^2 = 1$ can be explicitly verified. 
As in appendix \ref{APPA} it is useful to investigate the specific limit $\alpha \to 0$, $n_{1}\to 1 $ and $\beta\to 1$. In this case, defining 
$x_{1}= k \tau_{1}$ and $x_{2} = k \tau_{2}$ the exact form 
of $d_{\pm}(x_{1}, x_{2})$ can be written as:
\begin{eqnarray}
d_{+}(x_{1}, x_{2}) &=& \frac{e^{i x_{2}}}{16\, x_{1}^2 \,x_{2}^2}\biggl\{ e^{ - 2 i x_{2}} + e^{ 2 i x_{1}} [ 8 x_{2}^2 + 4 i x_{2} -1] [ 2 x_{1} ( x_{1} + i) - 1] \biggr\},
\label{dpex}\\
d_{+}(x_{1}, x_{2}) &=& \frac{e^{-3 i x_{2}}}{16\, x_{1}^2 \,x_{2}^2}\biggl\{ e^{  2 i x_{2}}[ 8 x_{2}^2 - 4 i x_{2}^2 -1] + e^{ 2 i x_{1}} [ 1 
- 2 x_{1} ( i + x_{1}) ]\biggr\}.
\label{dmex}
\end{eqnarray}
The expressions of Eqs. (\ref{dpex}) and (\ref{dmex}) can be expanded in order to derive 
approximate expressions of the transfer function of the relic graviton spectrum across the radiation-matter transition.
The exact results for the mixing coefficients have been used as a systematic cross-check: the power spectra 
and the energy density of the gravitons in the case of a dynamical refractive index are complicated functions 
of $\alpha$ which must anyway reduce to known expressions in the  $\alpha \to 0$ limit. This is not only 
true for the exact expressions but also for the corresponding approximated mixing coefficients 
computed in the limits $|k \tau_{1}| <1$ and $|k\tau_{2}| < 1$.
\end{appendix}

\end{document}